\documentclass[12pt,a4paper]{article}

\usepackage{epsfig}
\usepackage{amsmath,amsfonts,amssymb}
\usepackage{t1enc}
\usepackage{cite}

\providecommand{\openone}{\leavevmode\hbox{\small1\kern-3.8pt\normalsize1}}

\newcommand{\Gz}{\Gamma_0}
\newcommand{\Gpm}{\Gamma_{R,L}}
\newcommand{\Fi}{F_i}
\newcommand{\Fpm}{F_{R,L}}
\newcommand{\fp}{F_R}
\newcommand{\fm}{F_L}
\newcommand{\fz}{F_0}
\newcommand{\rhp}{\rho_R}
\newcommand{\rhm}{\rho_L}
\newcommand{\rhpm}{\rho_{R,L}}

\newcommand{\thlw}{\theta_{\ell}^*}

\newcommand{\hk}{\alpha}
\newcommand{\ptb}{p_T^{\bar b}}
\newcommand{\ptbisr}{p_T^{\bar b_\text{ISR}}}
\newcommand{\mwb}{m_{Wb}}

\newcommand{\afb}{A_\mathrm{FB}}
\newcommand{\Ap}{A_+}
\newcommand{\Am}{A_-}
\newcommand{\Apm}{A_\pm}

\newcommand{\rbl}{r_{bl}}
\newcommand{\rnl}{r_{\nu l}}

\newcommand{\vl}{V_L}
\newcommand{\vr}{V_R}
\newcommand{\gl}{g_L}
\newcommand{\gr}{g_R}

\newcommand{\glp}{f_{1L}}
\newcommand{\grp}{f_{1R}}
\newcommand{\hl}{f_{2L}}
\newcommand{\hr}{f_{2R}}

\begin{document}

\begin{center}
\begin{Large}
{\bf Single top quark production at LHC \\[2mm]
with anomalous $\boldsymbol{Wtb}$ couplings}
\end{Large}

\vspace{0.5cm}
J. A. Aguilar--Saavedra  \\[0.2cm] 
{\it Departamento de Física Teórica y del Cosmos and CAFPE, \\
Universidad de Granada, E-18071 Granada, Spain} \\[0.1cm]
\end{center}

\begin{abstract}
We investigate single top production in the presence of anomalous $Wtb$ couplings. We explicitly show that,
if these couplings arise from gauge invariant effective operators, the only relevant couplings for single top production and decay are the usual
$\gamma^\mu$ and $\sigma^{\mu \nu} q_\nu$ terms, where $q$ is the $W$ boson momentum. This happens even in the single top production processes where the $Wtb$ interaction involves off-shell top and/or bottom quarks.
With this parameterisation for the $Wtb$ vertex, we obtain expressions
for the dependence on anomalous couplings of the single top cross sections,
for (i) the $t$-channel process, performing a matching between $tj$ and $t \bar b j$ production, where $j$ is a light jet; (ii) $s$-channel $t \bar b$ production; (iii) associated $tW^-$ production, including the correction
from $tW^- \bar b$. We use these expressions to estimate, with a fast detector simulation, the simultaneous limits which the measurement of single top cross sections at LHC will set on $V_{tb}$ and possible anomalous couplings. Finally, a combination with top decay asymmetries and angular distributions is performed, showing how the limits can be improved when the latter are included in a global fit to $Wtb$ couplings.
\end{abstract}

\section{Introduction}

In addition to top pair production, which is the largest source of top quarks in hadron collisions, single top processes will be of great importance for the study of the top quark properties at the Large Hadron Collider (LHC)
 \cite{ Beneke:2000hk}. With a cross section about three times smaller than for $t \bar t$, single top quarks will also
be coupiously produced through the electroweak $Wtb$ interaction and, precisely because of this production mechanism, single top processes will provide essential information about the $Wtb$ vertex. Their cross sections are proportional to the size of this interaction, and thus from their measurement the value of the Cabibbo-Kobayashi-Maskawa (CKM) matrix element $V_{tb}$ will be obtained,
as well as bounds on possible anomalous couplings \cite{Boos:1999dd,Chen:2005vr}.
Moreover, single top quarks will be produced
with a high degree of polarisation \cite{Mahlon:1999gz}
(in contrast to QCD $t \bar t$ production),
which allows to build top spin asymmetries
of order $0.1-0.4$ in the leptonic decay $t \to Wb \to \ell \nu b$.
Such asymmetries complement the observables
measured in the $W$ rest frame \cite{AguilarSaavedra:2006fy}
in order to test anomalous $Wtb$ couplings with a high precision and in a model-independent way.

New physics beyond the Standard Model (SM) is expected to affect especially the top quark, and, in particular, it may modify its charged current interaction with its $\text{SU}(2)_L$ partner the bottom quark.
For on-shell $t$, $b$ and $W$, the most general $Wtb$ vertex involving terms
up to dimension five
can be written as \cite{AguilarSaavedra:2006fy}
\begin{eqnarray}
\mathcal{L}^\text{OS}_{Wtb} & = & - \frac{g}{\sqrt 2} \bar b \, \gamma^{\mu} \left( \vl
P_L + \vr P_R
\right) t\; W_\mu^- \nonumber \\
& & - \frac{g}{\sqrt 2} \bar b \, \frac{i \sigma^{\mu \nu} q_\nu}{M_W}
\left( \gl P_L + \gr P_R \right) t\; W_\mu^- + \mathrm{H.c.} \,,
\label{ec:1}
\end{eqnarray}
with $q \equiv p_t-p_b$ (being $p_t$ and $p_b$ the momenta of the top and $b$ quark, respectively, following the fermion flow), which equals the $W$ boson momentum.
Additional $\sigma^{\mu \nu} k_\nu$ and $k^\mu$ terms, where 
$k \equiv p_t+p_b$, can be absorbed into this Lagrangian using
Gordon identities. If the $W$ boson is on its mass shell or it couples to massless external
fermions we have $q^\mu \epsilon_\mu = 0$, where $\epsilon_\mu$ is the polarisation vector of the $W$ boson, so that terms proportional
to $q^\mu$  can be dropped from the effective vertex.
Within the SM, the only $Wtb$ interaction term at the tree level is given by the left-handed $\gamma^\mu$ term, with $\vl \equiv V_{tb} \simeq 1$.
The rest of couplings are called ``anomalous'' and vanish at the tree level, although they can be generated by radiative corrections.
They are not necessarily constants but rather ``form factors'', usually approximated by the constant term (as we will do in this work).
If we assume that CP is conserved in the $Wtb$ interaction then $V_{L,R}$ and
$g_{L,R}$ are real, and $\vl$ can be taken to be positive without loss of generality.

For off-shell top and/or bottom quarks the Lagrangian in Eq.~(\ref{ec:1})
is not the most general one, and in principle it should be extended with
$k^\mu$ and $\sigma^{\mu \nu} k_\nu$  terms.
Nevertheless, if $Wtb$ anomalous couplings arise from gauge invariant effective operators, single top production and decay can be described in full generality using the on-shell Lagrangian in Eq.~(\ref{ec:1}) for the $Wtb$ vertex,
even in the processes where the top and bottom quarks involved in the $Wtb$ interaction are far from their mass shell. This is a particular case of a more general set up, and it will be explicitly proved for the specific case of single top production. In particular, we will show that new physics contributions to the $Wtb$ vertex can be ``rewritten'' using Gordon identities 
into the form of Eq.~(\ref{ec:1}), even for $t$ and $b$ off-shell. The precise meaning of this ``rewriting'' will be clear in the next section: using an adequate parameterisation for the most general $Wtb$ vertex, including $k^\mu$ and $\sigma^{\mu \nu} k_\nu$ terms, we will see that off-shell effects of
$Wtb$ anomalous couplings in single top production identically cancel when summed to contributions from anomalous $gWtb$ quartic couplings, which are related to the former by gauge invariance. We refer the reader to section \ref{sec:eff}, where these issues are explained in detail. 

After showing that for single top production the general $Wtb$ vertex can be adequately described with the on-shell Lagrangian in Eq.~(\ref{ec:1}), we will obtain expressions for the single top cross sections in terms of the couplings
in Eq.~(\ref{ec:1}).
There are three single top (and antitop) production processes in hadron collisions: (i) the $t$-channel process, also denoted as $Wg$ fusion, which involves $tj$ and $t \bar b j$ production; (ii) $t \bar b$ production with a
$s$-channel $W$; (iii) associated $tW^-$ production, with a correction from the $t W^- \bar b$ process.
In the SM their cross sections are proportional to $\vl^2 (=V_{tb}^2)$.
For a general $Wtb$ interaction
the cross sections include additional terms involving anomalous $Wtb$ couplings and can be conveniently written, factorising the SM cross section (calculated for $\vl=1$), as
\begin{eqnarray}
\sigma & = & \sigma_\text{SM} \left( \vl^2 + \kappa^{\vr} \, \vr^2 + \kappa^{\vl \vr}\, \vl \vr + \kappa^{\gl} \, \gl^2 + \kappa^{\gr}\, \gr^2
 + \kappa^{\gl \gr}\, \gl \gr
 + \dots \right)  \,,
\label{ec:k}
\end{eqnarray}
where the $\kappa$ factors (9 numbers for the different combinations of couplings in $\mathcal{L}^\text{OS}_{Wtb}$) determine the dependence on anomalous couplings. These factors are in general different for $t$ and $\bar t$ production, and must be evaluated with a numerical integration of
the corresponding cross section. They depend on parton distribution functions (PDFs), the factorisation scale $Q^2$ and parameters such as the top and bottom quark masses.
These dependences translate into theoretical uncertainties which have to be considered when deriving limits on anomalous couplings from single top cross sections.

We will then take the next step and estimate the potential limits on $Wtb$
couplings from single top cross section measurements at LHC. These limits are not very stringent due to the experimental errors on the cross sections, which are expected to range between 13\% (for the $t$-channel process) and 21\% (for $t \bar b$ production). Finally, we will perform a combination with top decay observables, such as angular distributions and asymmetries, which complement the former and allow us to improve limits on $Wtb$ couplings significantly,
obtaining at the same time a measurement of $V_{tb}$ and limits on anomalous couplings.

The rest of this paper is organised as follows.
In section \ref{sec:eff} we derive the $Wtb$ couplings from gauge invariant effective operators and discuss the cancellation among off-shell contributions involving triple and quartic vertices. Readers mainly interested in the numerical results may skip this section. 
In sections \ref{sec:2}--\ref{sec:4} we study in turn the three single top production processes in the presence of anomalous couplings and derive the corresponding $\kappa$ factors for single top and antitop production.
In section \ref{sec:5} we present a first estimate for the limits on anomalous couplings which can be obtained from the measurement of single top cross sections, as well as with their combination with top decay angular distributions and asymmetries. Our conclusions are presented in section \ref{sec:6}. The explicit proof of the cancellation among off-shell and quartic contributions is given in appendix \ref{sec:a}. A brief review of the top decay observables used in the combination is given in appendix \ref{sec:b}.
The effect of the top decay is examined in appendix \ref{sec:c}.

\section{$Wtb$ anomalous couplings from effective operators}
\label{sec:eff}

We follow the notation of Ref.~\cite{Buchmuller:1985jz} for gauge invariant effective operators with slight normalisation changes and sign differences. For
reference, we summarise here the definitions needed in this section.
We denote by
\begin{equation}
q_L = \left(\! \begin{array}{c} t_L \\ b_L \end{array} \!\right) \;,\quad
 t_R \;,\quad b_R
\end{equation}
the weak interaction eigenstates for the third generation.
The covariant derivative is
\begin{equation}
D_\mu = \partial_\mu + i g_s \frac{\lambda^a}{2} G_\mu^a
+ i g \frac{\tau^I}{2} W_\mu^I + i g' Y B_\mu \,,
\end{equation}
where $\lambda^a$ are the Gell-Mann matrices with $a=1\dots 8$,
$\tau^I$ the Pauli matrices for $I=1,2,3$,
$Y$ is the hypercharge and $G_\mu^a$, $W_\mu^I$ and $B_\mu$ are the gauge fields for
$\mathrm{SU}(3)$, $\mathrm{SU}(2)_L$ and $\mathrm{U}(1)_Y$ respectively.
The charged
$W$ boson fields are
\begin{equation}
W^\pm = \frac{1}{\sqrt 2} \left( W^1 \mp i W^2 \right)
\end{equation}
and the field strength tensor for $\text{SU}(2)_L$ is
\begin{equation}
W_{\mu \nu}^I = \partial_\mu W_\nu^I - \partial_\nu W_\mu^I - g \epsilon_{IJK} W_\mu^J W_\nu^K \,.
\end{equation}
The SM Higgs doublet $\phi$ has vacuum expectation value
\begin{equation}
\langle \phi \rangle = \frac{1}{\sqrt 2}
\left(\! \begin{array}{c} 0 \\ v \end{array} \!\right) \,,
\end{equation}
with $v=246$ GeV, and we define $\tilde \phi = i \tau^2 \phi^*$.

The anomalous terms involving $\vr$, $\gr$ and $\gl$ in the on-shell Lagrangian of Eq.~(\ref{ec:1}) can arise from the effective operators
\begin{align}
& O_{\phi\phi} = i (\phi^\dagger i \tau^2 D_\mu \phi) \, (\bar t_R \gamma^\mu b_R)
& & \to \quad
\left[ \frac{g v^2}{2 \sqrt 2} \, \bar b_R \gamma^\mu t_R \, W_\mu^- \right]^\dagger
\,, \notag \\
& O_{uW} = (\bar q_L \sigma^{\mu \nu} \tau^I t_R) \,\tilde \phi \, W_{\mu \nu}^I
& &\to \quad
2 v \bar b_L \, i \sigma^{\mu \nu} q_\nu  t_R \, W_\mu^-
\,, \notag \\
& O_{dW} = (\bar q_L \sigma^{\mu \nu} \tau^I d_R) \,\phi\, W_{\mu \nu}^I
& &\to \quad
\left[ -2 v \bar b_R \, i \sigma^{\mu \nu} q_\nu  t_L \, W_\mu^- \right]^\dagger \,.
\label{ec:OW}
\end{align}
Contributions to $\vl$ can originate from
\begin{align}
& O_{\phi q}^{(3)} = i (\phi^\dagger D_\mu \tau^I \phi) \, (\bar q_L \gamma^\mu \tau^I q_L)
& & \to \quad
-\frac{g v^2}{\sqrt 2} \, \bar b_L \gamma^\mu t_L \, W_\mu^- \,.
\end{align}
Notice that these four operators do not yield quartic terms relevant for the amplitudes involved in single top production. Contributions to the $Wtb$ vertex involving the top and $b$ quark momenta
can be obtained from the operators
\begin{align}
& O_{Du} = (\bar q_L \,D_\mu t_R) \, D^\mu \,\tilde \phi
\,, \notag \\
& O_{\bar Du} = (D_\mu  \bar q_L \,t_R) \, D^\mu \,\tilde \phi
\,, \notag \\
& O_{Dd} = (\bar q_L \,D_\mu b_R) \, D^\mu \, \phi
\,, \notag \\
& O_{\bar Dd} = (D_\mu  \bar q_L \,b_R) \, D^\mu \, \phi
\,.
\end{align}
The opposite sign combinations
\begin{align}
& O_{Du}-O_{\bar Du} && \to \quad
\frac{gv}{2} \, \bar b_L k^\mu t_R \, W_\mu^-
- g g_s v \, \bar b_L \frac{\lambda^a}{2} g^{\mu \nu} t_R \, W_\mu^- \, G_\nu^a \,, \notag \\
& O_{Dd}-O_{\bar Dd} && \to \quad
\left[ \frac{gv}{2} \, \bar b_R k^\mu t_L \, W_\mu^-
- g g_s v \, \bar b_R \frac{\lambda^a}{2} g^{\mu \nu} t_L \, W_\mu^- \,
 G_\nu^a \right]^\dagger
\label{ec:OD}
\end{align}
give $k^\mu$ terms plus quartic $gWtb$ interactions, while the same sign combinations
\begin{align}
& O_{Du}+O_{\bar Du} && \to \quad
\frac{gv}{2} \, \bar b_L q^\mu t_R \, W_\mu^- \,, \notag \\
& O_{Dd}+O_{\bar Dd} && \to \quad
\left[ -\frac{gv}{2} \, \bar b_R q^\mu t_L \, W_\mu^- \right]^\dagger
\end{align}
give $q^\mu$ terms in the $Wtb$ vertex but not $gWtb$ ones, which cancel in the sums. Analogously, from the operators
\begin{align}
& O'_{Du} = i (\bar q_L \, \sigma^{\mu \nu}  D_\nu t_R) \, D_\mu \,\tilde \phi
\,, \notag \\
& O'_{\bar D u} = i (D_\nu  \bar q_L \, \sigma^{\mu \nu}  t_R) \, D_\mu \,\tilde \phi \,, \notag \\
& O'_{Dd} = i (\bar q_L \, \sigma^{\mu \nu}  D_\nu b_R) \, D_\mu \, \phi
\,, \notag \\
& O'_{\bar D d} = i (D_\nu  \bar q_L \, \sigma^{\mu \nu}  b_R) \, D_\mu \, \phi \,,
\end{align}
which are equivalent to the ones in Eq.~(\ref{ec:OD}) for $t$, $b$ on-shell \cite{Buchmuller:1985jz}, we can get contributions to $\sigma^{\mu \nu}$ terms. The combinations
\begin{align}
& O'_{Du}-O'_{\bar Du} && \to \quad
\frac{gv}{2} \, \bar b_L i \sigma^{\mu \nu} k_\nu t_R \, W_\mu^-
- g g_s v \, \bar b_L \frac{\lambda^a}{2} i \sigma^{\mu \nu} t_R \, W_\mu^- \, G_\nu^a \,, \notag \\
& O'_{Dd}-O'_{\bar Dd} && \to \quad
\left[ -\frac{gv}{2} \, \bar b_R i \sigma^{\mu \nu} k_\nu t_L \, W_\mu^-
+ g g_s v \, \bar b_R \frac{\lambda^a}{2} i \sigma^{\mu \nu} t_L \, W_\mu^- \,
 G_\nu^a \right]^\dagger
\label{ec:ODp}
\end{align}
yield $\sigma^{\mu \nu} k_\nu$ terms in the $Wtb$ vertex plus quartic $gWtb$ interactions, and the combinations
\begin{align}
& O'_{Du}+O'_{\bar Du} && \to \quad
\frac{gv}{2} \, \bar b_L i \sigma^{\mu \nu} q_\nu t_R \, W_\mu^- \,, \notag \\
& O'_{Dd}+O'_{\bar Dd} && \to \quad
\left[ \frac{gv}{2} \, \bar b_R i \sigma^{\mu \nu} q_\nu t_L \, W_\mu^- \right]^\dagger
\end{align}
give $\sigma^{\mu \nu} q_\nu$ terms (which can also be obtained from the operators $O_{uW}$ and $O_{dW}$ in Eq.~(\ref{ec:OW}), respectively) but not quartic $gWtb$ ones.

The most general $Wtb$ vertex for $t$, $b$ off-shell and $W$ on-shell or coupling to external masless fermions includes the Lagrangian in Eq.~(\ref{ec:1}) plus $\sigma^{\mu \nu} k_\nu$ and $k^\mu$ terms. But,
instead of simply adding the latter terms to the on-shell Lagrangian,
it is much more convenient to parameterise the general $Wtb$ vertex in terms of
$\mathcal{L}_{Wtb}^\text{OS}$ plus operators which vanish
when $t$ and $b$ are on their mass shell,
\begin{equation}
\mathcal{L}_{Wtb} = \mathcal{L}^\text{OS}_{Wtb} + \mathcal{O}_1
+ \mathcal{O}_2 \,,
\label{ec:2}
\end{equation}
with
\begin{eqnarray}
\mathcal{O}_1 & = & -\frac{g}{\sqrt 2 M_W} \, \bar b  \left[ i \sigma^{\mu \nu} k_\nu (\glp P_L + \grp P_R )
 - (m_b \glp - m_t \grp ) \gamma^\mu P_L
\right. \notag \\
& & \left. - (-m_t \glp + m_b \grp) \gamma^\mu P_R
- q^\mu (\glp P_L + \grp P_R ) 
 \right] t \; W_\mu^- + \mathrm{H.c.} \,, \notag \\[2mm]
\mathcal{O}_2 & = & -\frac{g}{\sqrt 2 M_W}  \, \bar b \left[ k^\mu (\hl P_L + \hr P_R ) 
- i \sigma^{\mu \nu} q_\nu (\hl P_L + \hr P_R ) 
\right. \notag \\
& & \left. -(m_b \hl + m_t \hr ) \gamma^\mu P_L 
-(m_t \hl + m_b \hr ) \gamma^\mu P_R \right] t \; W_\mu^-
+ \mathrm{H.c.}\,,
\label{ec:3}
\end{eqnarray}
being $f_{1L,1R}$, $f_{2L,2R}$ arbitrary constants.
We point out that $\mathcal{O}_1$, $\mathcal{O}_2$ are just the
Gordon identities for the $\sigma^{\mu \nu} k_\nu$ and $k^\mu$ terms
properly normalised. 
This parameterisation is completely general for a CP-conserving $Wtb$ vertex. From Eqs.~(\ref{ec:OD}), (\ref{ec:ODp}) we observe that corresponding to the off-shell operators $\mathcal{O}_{1,2}$, the anomalous $gWtb$ interactions
\begin{eqnarray}
\mathcal{O}_1^{(4)} & = & \frac{\sqrt 2 g g_s}{M_W} \, \bar b \frac{\lambda^a}{2} \, i \sigma^{\mu \nu} \left( \glp P_L + \grp P_R \right)
t \, W_\mu^- G_\nu^a + \mathrm{H.c.} \,, \notag \\
\mathcal{O}_2^{(4)} & = & \frac{\sqrt 2 g g_s}{M_W} \, \bar b \frac{\lambda^a}{2} \, g^{\mu \nu} \left( \hl P_L + \hr P_R \right)
t \, W_\mu^- G_\nu^a + \mathrm{H.c.} 
\label{ec:O4}
\end{eqnarray}
have to be introduced if the former arise from gauge invariant effective operators.

\begin{figure}[ht]
\begin{center}
\begin{tabular}{cccccc}
\raisebox{-1.5cm}{\epsfig{file=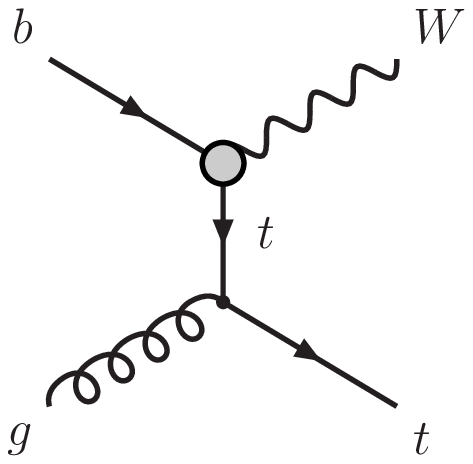,height=3cm,clip=}} &
{\Large $+$} & 
\raisebox{-1.5cm}{\epsfig{file=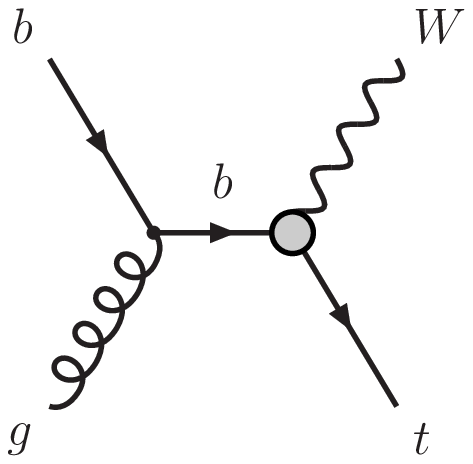,height=3cm,clip=}} &
{\Large $+$} &
\raisebox{-1.5cm}{\epsfig{file=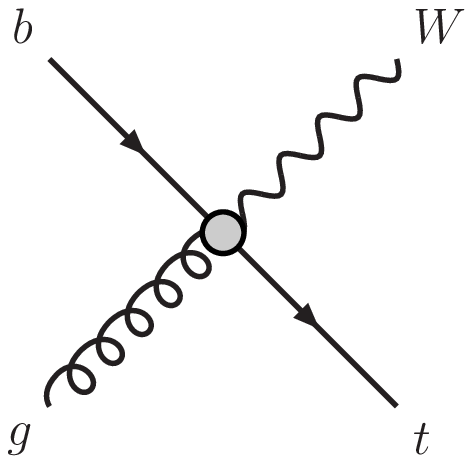,height=3cm,clip=}} &
{\Large $= \quad 0$}
\end{tabular}
\end{center}
\caption{Cancellation between contributions to the $gb \to tW$ amplitude.
The gray circles in the diagrams stand for: (i) a $Wtb$ interaction through $\mathcal{O}_1$ in the first and second diagrams, and a $gWtb$ interaction through $\mathcal{O}_1^{(4)}$ in the third one; or
(ii) the same involving $\mathcal{O}_2$ and  $\mathcal{O}_2^{(4)}$.}
\label{fig:cancel}
\end{figure}

A very important consequence of gauge invariance is the cancellation among the contributions of $\mathcal{O}_{1,2}$ and $\mathcal{O}_{1,2}^{(4)}$ to the amplitudes. For the case of $gb \to tW^-$ this is simbolically depicted in
Fig.~\ref{fig:cancel}. 
We emphasise that the cancellation takes place when {\em all} the
terms in $\mathcal{O}_1$ (or $\mathcal{O}_2$) written in Eqs.~(\ref{ec:3}) are included, not only the $\sigma^{\mu \nu} k_\nu$ (or $k^\mu$) ones.
This clarification is important because only the $\sigma^{\mu \nu} k_\nu$ and $k^\mu$ terms in $\mathcal{O}_{1,2}$ have associated quartic vertices 
in Eq.~(\ref{ec:O4}), but choosing the parameterisation in
Eq.~(\ref{ec:3}) makes the cancellation apparent, decreasing the number of couplings relevant for single top production from 8 to 4. The on-shell Lagrangian in Eq.~(\ref{ec:1})
can then be used in full generality for the study of $tW^-$ production \cite{Najafabadi:2008pb}, as well as in the $t \bar b j$ and $tW^- \bar b$ processes, for which the $Wtb$ vertex involves off-shell $t$ and/or $b$ but the effects of $\mathcal{O}_{1,2}$ and $\mathcal{O}_{1,2}^{(4)}$ cancel.
(For $tj$ and $t \bar b$ the top and bottom quarks are already on-shell.)
The explicit proof of the cancellations is given in appendix \ref{sec:a}
for the $gb \to tW^-$ and $gg \to tW^- \bar b$ processes.
It is interesting to point out that
in the latter case it implies that off-shell operators do not influence the top decay, and for $t \bar t$ production with decay of both quarks, {\em i.e.} $gg \to W^+ b W^- b$, the cancellation is expected to take place as well, although it is not proved explicitly.
For $t \bar b j$ the diagrams can be related to $tW^-$ production by crossing and the addition of a fermion line to the external $W$ boson, and the proof is the same.

A consequence of the gauge cancellation is that, as we have mentioned in the introduction, we can actually use Gordon identities to rewrite new physics contributions into the form of Eq.~(\ref{ec:1}), even for $t$ and $b$ off-shell. Let us assume that
some new physics at a high scale gives a contribution to the $Wtb$
vertex of the form
\begin{equation}
\Delta \mathcal{L} = -\frac{v}{\Lambda^2} \, \bar b \, k^\mu (c_L P_L + c_R P_R) \,
t \, W_\mu^- + \mathrm{H.c.} \,,
\end{equation}
with $c_L$, $c_R$ constants, plus additional triple and quartic contributions implied by gauge invariance. We can also write $\Delta \mathcal{L}$ as
\begin{equation}
\Delta \mathcal{L} = \Delta \mathcal{L}_1 + \Delta \mathcal{L}_2 \,,
\end{equation}
with
\begin{eqnarray}
\Delta \mathcal{L}_1 & = & - \frac{v}{\Lambda^2} \, \bar b 
\left[ k^\mu (c_L P_L + c_R P_R) -i \sigma^{\mu \nu} q_\nu (c_L P_L + c_R P_R)
\right. \nonumber \\[1mm]
& & \left. -(m_b c_L + m_t c_R ) \gamma^\mu P_L 
-(m_t c_L + m_b c_R ) \gamma^\mu P_R \right]
t \, W_\mu^- + \mathrm{H.c.} \,, \nonumber \\[1mm]
\Delta \mathcal{L}_2 & = & - \frac{v}{\Lambda^2} \, \bar b 
\left[ i \sigma^{\mu \nu} q_\nu (c_L P_L + c_R P_R)
+ (m_b c_L + m_t c_R ) \gamma^\mu P_L \right. \nonumber \\[1mm]
& & \left. + (m_t c_L + m_b c_R ) \gamma^\mu P_R \right] t \, W_\mu^- + \mathrm{H.c.}
\end{eqnarray}
Because of the cancellation mentioned, the contribution of $\Delta \mathcal{L}_1$
(which has the same form as $\mathcal{O}_2$, with $g f_{1L,2L} = (2 M_W v/\Lambda^2)
\, c_{L,R}$) to single top production and decay vanishes, thus we can effectively make the replacement
\begin{equation}
\Delta \mathcal{L} \to \Delta \mathcal{L}_2 \,,
\end{equation}
which amounts to using the Gordon identities on $\Delta \mathcal{L}$ independently of whether the top and bottom quarks are on-shell or not.
Obviously, the same statement applies to $\sigma^{\mu \nu} k_\nu$-type contributions
as well.

\section{The $t$-channel process}
\label{sec:2}

\begin{figure}[ht]
\begin{center}
\begin{tabular}{ccccc}
\epsfig{file=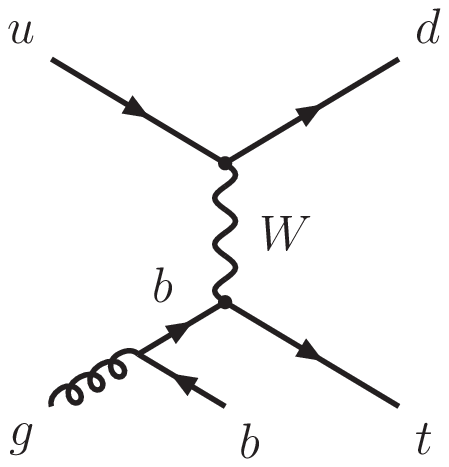,height=3cm,clip=} & \quad &
\epsfig{file=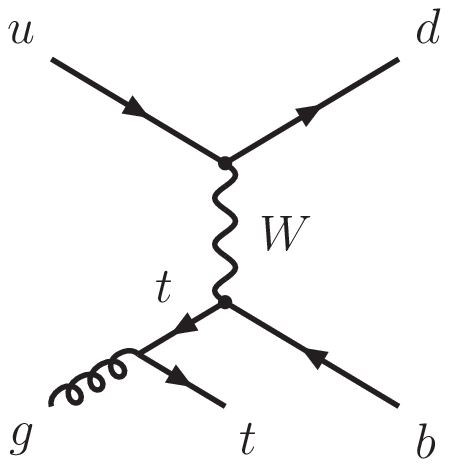,height=3cm,clip=} & \quad & 
\epsfig{file=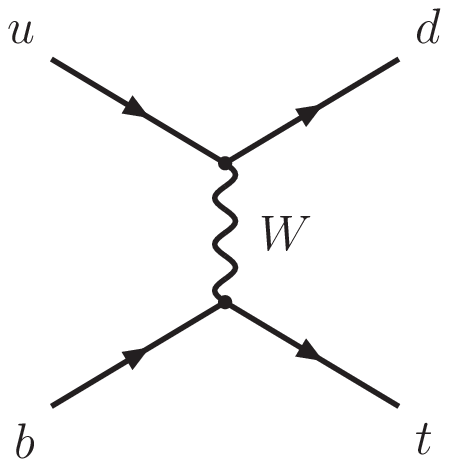,height=3cm,clip=}  \\
(a) & & (b) & & (c)
\end{tabular}
\caption{Sample Feynman diagrams for single top production in the
$t$-channel process. Additional diagrams are obtained by crossing the light quark fermion line, and/or replacing $(u,d)$ by $(c,s)$. The diagrams for antitop production are the charge conjugate ones.}
\label{fig:diag1}
\end{center}
\end{figure}

\begin{figure}[htb]
\begin{center}
\begin{tabular}{c}
\epsfig{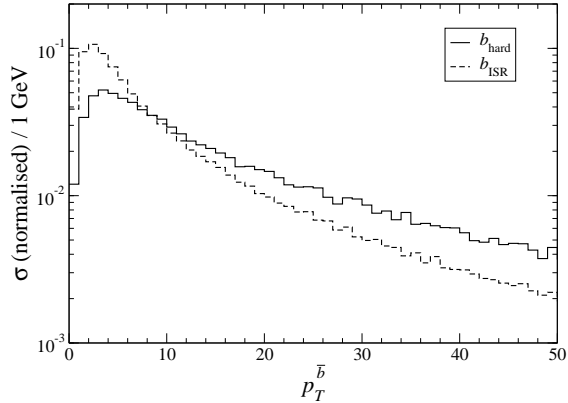}
\end{tabular}
\caption{Normalised transverse momentum distribution for the $\bar b$ quark in the
$t$-channel process, when it is generated from the hard process or from ISR
in the parton-shower Monte Carlo.}
\label{fig:ptbisr}
\end{center}
\end{figure}

This process involves the scattering of a light quark and a gluon from the proton sea, as shown in Fig.~\ref{fig:diag1} (a,b).
(We will often refer to the $t \bar b j$
and $tj$ processes generically, understanding that the same applies to the charge conjugate processes $\bar t bj$ and $\bar tj$. When there is some difference we will note it explicitly.) Diagram (a) where the gluon splits into a $b \bar b$ pair is the dominant one in the SM, but diagram (b) has to be included as well in order to form a gauge invariant set.
For low transverse momentum $\ptb$ of the $\bar b$ quark,
there are large logarithmic corrections to the cross section from the internal $b$ propagator, which can be resummed by introducing a $b$ quark
PDF in the proton
\cite{Willenbrock:1986cr} and describing the process as $2 \to 2$ body, as shown in  diagram (c).
The extra $\bar b$ quark is generated in this case from initial state radiation (ISR) in a parton shower Monte Carlo. When merging
the $2 \to 2$ and $2 \to 3$ processes, one has to be careful to avoid double counting in the low $\ptb$ region. One possibility
is to perform a matching as suggested in Ref.~\cite{Stelzer:1998ni}, based on the $\ptb$ distribution shown in Fig.~\ref{fig:ptbisr}.
For the $t\bar b j$ process it is required that the $\bar b$ quark has transverse momentum above certain value, $\ptb > p_T^\text{cut}$, while for the $tj$ process a veto
$\ptbisr < p_T^\text{cut}$ is imposed in the parton shower Monte Carlo.
The cross section for the low-$\ptb$ region is determined from the next-to-leading order (NLO) value and the cross section above the cut, for which the perturbative calculation is reliable. This amounts to normalise the cross section in the low-$\ptb$ region with a $K$ factor,
\begin{equation}
K \sigma(tj, \ptbisr < p_T^\text{cut}) =
\sigma_\text{NLO} - \sigma(t \bar b j, \ptb > p_T^\text{cut}) \,.
\label{ec:tjmatch}
\end{equation}
As it has been shown in Ref.~\cite{Boos:2006af}, the choice $p_T^\text{cut} = 10$ GeV
leads to a smooth transition between the high-$\ptb$ region from diagrams (a,b) and the low-$\ptb$ region, described by diagram (c) plus an additional $\bar b_\text{ISR}$ quark from ISR. Furthermore, it gives very good agreement for the kinematical distributions
of various variables with the full QCD NLO calculation, as implemented in the generators
{\tt ZTOP} \cite{Sullivan:2004ie} and {\tt MCFM} \cite{Campbell:2004ch}.
This is important in order to obtain an accurate prediction for the $\kappa$ coefficients corresponding to $\sigma^{\mu \nu}$ couplings, in which the vertex involves a momentum factor.
This matching prescription also has the advantage that it avoids introducing negative weight events, as it happens with procedures involving a subtraction term \cite{Barnett:1987jw,Olness:1987ep} and can easily be implemented in a Monte Carlo generator. Electroweak corrections are small, only a $-1.5\%$ \cite{Beccaria:2008av}, and do not modify the kinematics.

For each $tj$, $t \bar b j$ sub-process $bu \to td$,
$gu \to t \bar b d$, etc. (some of the amplitudes are related by crossing symmetry) the squared matrix elements are calculated analytically with
{\tt FORM} \cite{Vermaseren:2000nd}, taking the $Wtb$ interaction in
Eq.~(\ref{ec:1}) and keeping $m_b$ non-vanishing.
Then, the contributions to the cross section of the different
products of anomalous couplings in Eq.~(\ref{ec:k}) are obtained by identifying in the squared matrix element
the term multiplying each combination of couplings, and then integrating these terms separately over phase space and PDFs.
Obviously, the cross section is the sum of all terms, and the $\kappa$ coefficients
in Eq.~(\ref{ec:k}) are obtained by dividing by the $\vl^2$ term.
A cross-check of all matrix elements is performed calculating them numerically with {\tt HELAS} \cite{helas}, extended to include the non-renormalisable $\sigma^{\mu \nu}$ and $k^\mu$ couplings in Eqs.~(\ref{ec:1}) and (\ref{ec:3}). Both the analytical and numerical matrix elements agree within several digits. For the calculation of the cross sections
we use CTEQ6M \cite{Pumplin:2002vw} PDFs with the scales advocated in Ref. \cite{Stelzer:1997ns}: $Q^2 = q^2$ for the light quark, $Q^2 = q^2+m_t^2$ for the initial $b$ and
$Q^2=(\ptb)^2+m_b^2$ for the gluon ($q$ is the momentum of the $t$-channel $W$ boson). We take $m_t = 175$ GeV, $M_W = 80.39$ GeV,
$m_b = 4.8$ GeV, and use the NLO cross sections $\sigma_\text{NLO}(t) = 155.9^{+5.8}_{-6.3}$ pb, $\sigma_\text{NLO}(\bar t) = 90.7^{+3.4}_{-3.7}$ pb \cite{Sullivan:2004ie} (the uncertainties have been rescaled to $\Delta m_t = 1.8$
GeV).

The cut $\ptb > 10$ GeV in the $t \bar b j$ process can readily be performed at the generator level, obtaining cross sections $\sigma(t \bar bj) = 80.9$ pb,
$\sigma(\bar t bj) = 48.3$ pb.
The veto $\ptbisr < 10$ GeV for the $tj$ process is more involved. It is implemented by linking {\tt Pythia} 6.4 \cite{Sjostrand:2006za}
to the Monte Carlo integration program. During the numerical integration of the cross section, each weighted event is feeded into {\tt Pythia}, which adds the ISR. Subsequently, the event is accepted (in which case it is considered for the integration) or rejected.
The effect of the $\ptbisr < 10$ veto on the $\kappa$ coefficients amounts to a few percent, thus the systematic uncertainty originating from ISR modeling in {\tt Pythia} is expected to be small. 
The results for the most relevant coefficients in $tj$, $\bar t j$, $t \bar b j$
and $\bar t bj$ are shown in Tables \ref{tab:tj} and \ref{tab:tbj}, omitting for brevity the coefficients which are smaller than 0.1 for both $t$ and $\bar t$ production.
In each block, the first column corresponds to the interval of variation when
CTEQ6M or MRST 2006 \cite{Martin:2007bv} PDFs are used. The second column is the uncertainty from the choice of factorisacion scale, and the third and fourth columns 
represent the uncertainties corresponding to $\Delta m_t = 1.8$ GeV and
$\Delta m_b = 220$ MeV, respectively.  Monte Carlo statistical uncertainties are of order $10^{-3}$ in most cases, and the same random seeds are used in all samples in order to reduce the statistical fluctuations among them.
\begin{table}[htb]
\begin{footnotesize}
\begin{center}
\begin{tabular}{cclll|clll}
           & \multicolumn{4}{c}{$t j$} & \multicolumn{4}{c}{$\bar t j$}\\[1mm]
           & $\kappa$         & $\Delta Q$           & $\Delta m_t$         & $\Delta m_b$
           & $\kappa$         & $\Delta Q$           & $\Delta m_t$         & $\Delta m_b$         \\[2mm]
$\vr^2$    & $0.916-0.923$    & $^{+0.}_{-0.}$       & $^{+0.}_{-0.}$       & $^{+0.}_{-0.}$      
           & $1.082-1.084$    & $^{+0.}_{-0.}$       & $^{+0.}_{-0.}$       & $^{+0.}_{-0.}$       \\[1mm]
$\gl^2$    & $1.75-1.79$      & $^{+0.044}_{-0.038}$ & $^{+0.007}_{-0.035}$ & $^{+0}_{-0.027}$ 
           & $2.16-2.17$      & $^{+0.035}_{-0.022}$ & $^{+0.014}_{-0.032}$ & $^{+0.}_{-0.}$       \\[1mm]
$\gr^2$    & $2.18$           & $^{+0.042}_{-0.033}$ & $^{+0.014}_{-0.034}$ & $^{+0.}_{-0.022}$
           & $1.75-1.77$      & $^{+0.042}_{-0.033}$ & $^{+0.007}_{-0.033}$ & $^{+0.}_{-0.025}$ \\[1mm]
$\vl \gr$  & $-(0.348-0.365)$ & $^{+0.007}_{-0.011}$ & $^{+0.}_{-0.}$       & $^{+0.}_{-0.}$      
           & $-(0.038-0.040)$ & $^{+0.010}_{-0.009}$ & $^{+0.}_{-0.}$       & $^{+0.}_{-0.}$       \\[1mm]
$\vr \gl$  & $-(0.006-0.008)$ & $^{+0.006}_{-0.005}$ & $^{+0.}_{-0.}$       & $^{+0.}_{-0.}$      
           & $-(0.399-0.408)$ & $^{+0.}_{-0.008}$    & $^{+0.}_{-0.}$       & $^{+0.}_{-0.}$
\end{tabular}
\end{center}
\end{footnotesize}
\caption{Representative $\kappa$ factors for the $tj$ and $\bar t j$ processes and their uncertainties, explained in the text. Errors smaller than $0.005$ are omitted.}
\label{tab:tj}
\end{table}
\begin{table}[htb]
\begin{footnotesize}
\begin{center}
\begin{tabular}{cclll|clll}
           & \multicolumn{4}{c}{$t \bar b j$} & \multicolumn{4}{c}{$\bar t b j$}\\[1mm]
           & $\kappa$         & $\Delta Q$           & $\Delta m_t$         & $\Delta m_b$
           & $\kappa$         & $\Delta Q$           & $\Delta m_t$         & $\Delta m_b$         \\[2mm]
$\vr^2$    & $0.927-0.932$    & $^{+0.005}_{-0.}$    & $^{+0.}_{-0.}$       & $^{+0.}_{-0.}$      
           & $1.068-1.069$    & $^{+0.}_{-0.005}$    & $^{+0.}_{-0.}$       & $^{+0.}_{-0.}$       \\[1mm]
$\vl \vr$  & $-0.117$         & $^{+0.}_{-0.}$       & $^{+0.}_{-0.}$       & $^{+0.005}_{-0.005}$ 
           & $-0.126$         & $^{+0.}_{-0.}$       & $^{+0.}_{-0.}$       & $^{+0.006}_{-0.006}$ \\[1mm]
$\gl^2$    & $1.96-2.01$      & $^{+0.070}_{-0.056}$ & $^{+0.005}_{-0.005}$ & $^{+0.}_{-0.}$       
           & $2.98-3.00$      & $^{+0.040}_{-0.040}$ & $^{+0.014}_{-0.014}$ & $^{+0.}_{-0.}$       \\[1mm]
$\gr^2$    & $2.97-2.98$      & $^{+0.056}_{-0.043}$ & $^{+0.013}_{-0.013}$ & $^{+0.}_{-0.}$      
           & $2.08-2.11$      & $^{+0.056}_{-0.045}$ & $^{+0.006}_{-0.007}$ & $^{+0.}_{-0.}$       \\[1mm]
$\vl \gr$  & $-(0.539-0.550)$ & $^{+0.012}_{-0.010}$ & $^{+0.}_{-0.}$       & $^{+0.}_{-0.}$      
           & $-(0.169-0.172)$ & $^{+0.010}_{-0.010}$ & $^{+0.014}_{-0.013}$ & $^{+0.}_{-0.}$       \\[1mm]
$\vr \gl$  & $-(0.121-0.134)$ & $^{+0.009}_{-0.011}$ & $^{+0.}_{-0.}$       & $^{+0.}_{-0.}$      
           & $-(0.567-0.571)$ & $^{+0.014}_{-0.013}$ & $^{+0.}_{-0.}$       & $^{+0.}_{-0.}$
\end{tabular}
\end{center}
\end{footnotesize}
\caption{Representative $\kappa$ factors for the $t \bar b j$ and $\bar t b j$ processes and their uncertainties, explained in the text. Errors smaller than $0.005$ are omitted.}
\label{tab:tbj}
\end{table}
The most salient features of the results in Tables \ref{tab:tj} and \ref{tab:tbj}
are:
\begin{itemize}
\item[(i)] The coefficients of the $\vr^2$ terms are not equal to unity, in contrast with the other single top processes studied in the next two sections.
This is easy to understand with the examination of the squared matrix element for the $bu \to td$ and $u \bar d \to t \bar b$ processes, related by crossing symmetry. In both cases the coefficient of $\vl^2$ is proportional to $(p_u \cdot p_b) (p_d \cdot p_t)$ (see Fig.~\ref{fig:diag1}), while the coefficient of $\vr^2$
is proportional to $(p_u \cdot p_t) (p_d \cdot p_b)$.
In the $s$-channel process $u \bar d \to t \bar b$, integration in phase space renders the coefficients of $\vl^2$ and $\vr^2$ equal, but this is not the case for the $t$-channel process $bu \to td$. In the latter, the two coefficients are different even in the case of all fermions massless.
\item[(ii)] The coefficient of the $\vr^2$ term is different for single top and single antitop production, but the differences cancel to a large extent in the total cross section. This property makes the ratio $R(\bar t/t)=\sigma(\bar t)/\sigma(t)$ more sensitive to a $\vr$ component than the total cross section itself.
A purely left-handed interaction yields a total (top plus antitop) cross-section 
of 246 pb, while a purely right-handed interaction gives a total of 241 pb.  Even in this extreme case, the difference is only of $1.9\%$, too small to be observed given the experimental uncertainties involved (see section \ref{sec:5}). However,
for a left-handed interaction $R(\bar t/t) = 0.58$, while for a right-handed interaction $R(\bar t/t) = 0.68$, and this 17\% difference should be visible.
Nevertheless, its sensitivity to $\vr$ is superseeded by top decay angular distributions \cite{AguilarSaavedra:2007rs,Hubaut:2005er} (see appendix \ref{sec:b}.).\footnote{A $\vr$ term is indirectly constrained by the measured rate of $b \to s \gamma$
\cite{Larios:1999au,Burdman:1999fw,Grzadkowski:2008mf,delAguila:2008iz}
but, if our aim is to obtain direct and model-independent measurements of the $Wtb$ interaction we cannot rely on such indirect measurements which, on the other hand, are much more restrictive.}
\item[(iii)] The coefficients of the $\gl^2$ and $\gr^2$ terms are larger than unity, as expected from the $q_\nu$ enhancement factor in the interaction. Moreover, they are larger for $t \bar b j$ than for $tj$, due to the larger energy involved when
$\ptb$ is larger (hence the importance of describing correctly the tail of the
$\ptb$ distribution).
\item[(iv)] Interferences among anomalous couplings are important, and in some cases the corresponding terms have coefficients of order unity. This implies that taking only one nonzero anomalous coupling at a time is by far a too simplistic assumption, and the possible cancellations among anomalous couplings have to be explored and constrained using as much information as possible, from single top production cross sections as well as from top decay asymmetries.
\end{itemize}

We conclude this section by comparing these results with the ones obtained using
an alternative procedure to remove double counting, which is to subtract from the cross section a term involving
the first order logarithmic corrections included in the $2 \to 3$ process
\cite{Barnett:1987jw,Olness:1987ep}.
This term $\sigma_\text{sub}$ is calculated by replacing 
in the calculation of the $bq \to tj$ cross sections
the $b$ quark PDF by
\begin{equation}
\tilde f_b(x,Q) = \frac{\alpha_s(Q)}{2 \pi} \log \frac{Q^2}{m_b^2}
\int_x^1 \frac{dz}{z} \left[ \frac{z^2+(1-z)^2}{2} \right] f_g(x/z,Q) \,,
\label{ec:bpdf}
\end{equation}
where $f_g$ is the gluon PDF. For convenience, the double counting term can be subtracted from the $tj$ contribution.
The cross sections obtained can be read in Table~\ref{tab:tjsub}. In order to compare with the results obtained using
the matching prescription we separate the $tj$ and $t \bar b j$ cross sections in two regions, depending on whether $\ptb$ is larger or smaller than 10 GeV. The total cross section is larger than in previous case, where we normalised cross sections to the NLO value.
Notice that with the subtraction method a 20\% of the cross section for $\ptb > 10$ GeV is given by the $2 \to 2$  process $tj$ plus ISR, which has a softer spectrum.
The $\kappa$ factors for each region can be calculated by summing the $tj$ (minus the subtraction term) and $t \bar bj$ contributions for each coupling and dividing the result by
the $\vl^2$ term. The most important $\kappa$ factors are listed in Table~\ref{tab:ksub},
calculated with this method and with the matching at 10 GeV. We also include the values obtained using only the $t \bar b j$ process, for both the low and high $\ptb$ regions.

\begin{table}[ht]
\begin{center}
\begin{tabular}{cccc}
   & $\ptb < 10$ GeV & $\ptb > 10$ GeV & Total \\
$\sigma(tj)-\sigma_\text{sub}(tj)$
   & 28.3            & 16.4            & 44.6 \\
$\sigma(t \bar bj)$
   & 50.4            & 80.9            & 131.3 \\
$\sigma(\bar tj)-\sigma_\text{sub}(\bar tj)$
   & 18.1            & 10.8            & 28.9 \\
$\sigma(\bar t bj)$
   & 31.6            & 48.3            & 79.9
\end{tabular}
\end{center}
\caption{Cross sections (in pb) for the $t$-channel process when the subtraction procedure is applied.}
\label{tab:tjsub}
\end{table}

\begin{table}[ht]
\begin{center}
\begin{tabular}{ccccccc}
& \multicolumn{6}{c}{Single top}  \\
& \multicolumn{3}{c}{$\ptb < 10$ GeV} & \multicolumn{3}{c}{$\ptb > 10$ GeV} \\
           & M     & S     & $t \bar b j$ & M     & S     & $t \bar b j$ \\
$\vr^2$    & 0.915 & 0.921 & 0.916        & 0.927 & 0.923 & 0.927        \\
$\gl^2$    & 1.75  & 1.61  & 1.60         & 1.96  & 1.98  & 1.96         \\
$\gr^2$    & 2.18  & 1.99  & 2.01         & 2.97  & 2.90  & 2.97         \\[5mm]
\hline \\
& \multicolumn{6}{c}{Single antitop} \\
& \multicolumn{3}{c}{$\ptb < 10$ GeV} & \multicolumn{3}{c}{$\ptb > 10$ GeV} \\
           & M     & S     & $t \bar b j$ & M     & S     & $t \bar b j$ \\
$\vr^2$    & 1.084 & 1.075 & 1.081        & 1.068 & 1.073 & 1.068        \\
$\gl^2$    & 2.16  & 2.03  & 2.11         & 2.98  & 2.90  & 2.98          \\
$\gr^2$    & 1.75  & 1.66  & 1.71         & 2.08  & 2.08  & 2.08 
\end{tabular}
\end{center}
\caption{Comparison of some $\kappa$ factors calculated with the matching (M) and subtraction (S) prescriptions, and using only the $t \bar b j$ process}
\label{tab:ksub}
\end{table}

For $\kappa^{\vr}$ the agreement is very good, better than 1\%. For $\kappa^{\gl}$ and
$\kappa^{\gr}$ the values obtained with the subtraction mathod are generally smaller. The agreement
for $\ptb > 10$ GeV, where the dependence on anomalous couplings is stronger, is better than 3\%, of the same order of
the theoretical uncertainties quoted in Table~\ref{tab:tbj}. The differences
for $\ptb < 10$ GeV are more significant but always below 10\%.
In summary, both methods for removing double counting give results not very different, and similar to the ones obtained just using the $t \bar b j$ process.
The results obtained with the matching are believed to be more accurate, for the reasons explained above.

\section{$\boldsymbol{t \bar b}$ production}
\label{sec:3}

\begin{figure}[ht]
\begin{center}
\begin{tabular}{c}
\epsfig{file=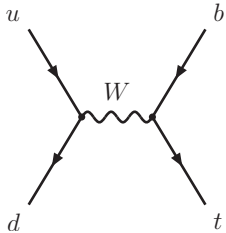,height=3cm,clip=}
\end{tabular}
\caption{Feynman diagram for single top production in the
$t \bar b$ process. An additional diagram is obtained replacing $(u,d)$ by $(c,s)$. For antitop production the diagrams are the charge conjugate ones.}
\label{fig:diag2}
\end{center}
\end{figure}

This process is mediated by $s$-channel Feynman diagrams as the one depicted in
Fig.~\ref{fig:diag2}. In Ref.~\cite{Sullivan:2004ie} it has been shown that NLO corrections do not significantly modify the kinematics and can be accounted 
for by a $K$ factor. Thus, the effect of anomalous $Wtb$ couplings is expected to be well approximated by including this $K$ factor in the tree-level cross section involving anomalous couplings. The NLO cross sections are
$\sigma_\text{NLO}(t) = 6.56 \pm 0.38$ pb,
$\sigma_\text{NLO}(\bar t) = 4.09 \pm 0.24$ pb~\cite{Sullivan:2004ie}.
Partial higher-order results are also known \cite{Kidonakis:2007ej}, and increase the $t \bar b$ cross section by about $\sim 10\%$. Their impact in the $t \bar b$ final state experimentally analysed, consisting of $t \bar b$ production plus contamination from $tj$ and $t \bar b j$ (see section \ref{sec:5}) is of only 3\%, much smaller than the experimental uncertainty which is around $20\%$. Anyway, they can be included with a $K$ factor.

In our calculations we use CTEQ6M PDFs with $Q = \sqrt s$, obtaining
$\sigma(t \bar b) = 4.62 $ pb, $\sigma(\bar t b) = 2.88 $ pb. The calculation of the $\kappa$ factors and their uncertainties proceeds in the same way as for $tj$ and $t \bar b j$ in the previous section. Results are shown in Table~\ref{tab:tb},
including uncertainties from PDFs, factorisation scale, the top and the $b$ quark masses.
We point out that the factors multiplying $\vl^2$ and $\vr^2$ in the cross section are equal, then $\kappa^{\vr}$ equals unity. We also have  $\kappa^{\gl} = \kappa^{\gr}$, $\kappa^{\vl \gl} = \kappa^{\vr \gr}$,
 $\kappa^{\vl \gr} = \kappa^{\vr \gl}$. From Table~\ref{tab:tb} we observe that:
\begin{itemize}
\item[(i)] The $\kappa$ factors of $\gl^2$ and $\gr^2$ are a factor of four larger than for the $t$-channel process, because in $t \bar b$ production the $s$-channel $W$ boson carries a larger momentum, and so the $q_\nu$ factor in the $\sigma^{\mu \nu}$ vertex gives a larger enhancement.
\item[(ii)] For $t \bar b$ and $\bar t b$ production the factors are very similar, although not equal (the difference is not due to Monte Carlo statistics, which is very high). Then, the measurement of the ratio $\sigma(\bar t b)/\sigma(t \bar b)$ is not as useful as in the $t$-channel process.
\item[(iii)] Interferences among couplings are again important, in particular between $\vl$ and $\gr$, and between $\vr$ and $\gl$.
\end{itemize}
Finally, it must be remarked that $t \bar b$ production, with a cross section much smaller than the $t$-channel process, has a large contamination from the latter in a real experiment. This is taken into account in the limits 
presented in section~\ref{sec:5}.

\begin{table}[t]
\begin{footnotesize}
\begin{center}
\begin{tabular}{cclll|clll}
                  & \multicolumn{4}{c}{$t \bar b$} & \multicolumn{4}{c}{$\bar t b$}\\[1mm]
                  & $\kappa$      & $\Delta Q$           & $\Delta m_t$         & $\Delta m_b$
                  & $\kappa$      & $\Delta Q$           & $\Delta m_t$         & $\Delta m_b$         \\[2mm]
$\vr^2$           & $1$           & --                   & --                   & --
                  & $1$           & --                   & --                   & --  \\[1mm]
$\vl \vr$         & $0.121$       & $^{+0.}_{-0.}$       & $^{+0.}_{-0.}$       & $^{+0.005}_{-0.005}$
                  & $0.127$       & $^{+0.}_{-0.}$       & $^{+0.}_{-0.}$       & $^{+0.006}_{-0.006}$ \\[1mm]
$\gl^2,\gr^2$     & $13.06-13.10$ & $^{+0.25}_{-0.21}$   & $^{+0.26}_{-0.26}$   & $^{+0.}_{-0.}$ 
                  & $12.22-12.28$ & $^{+0.21}_{-0.18}$   & $^{+0.25}_{-0.24}$   & $^{+0.}_{-0.}$       \\[1mm]
$\gl \gr$         & $1.23$        & $^{+0.007}_{-0.008}$ & $^{+0.012}_{-0.012}$ & $^{+0.055}_{-0.055}$
                  & $1.25$        & $^{+0.008}_{-0.009}$ & $^{+0.013}_{-0.013}$ & $^{+0.056}_{-0.056}$ \\[1mm]
$\vl\gl,\vr\gr$   & $-0.415$      & $^{+0.}_{-0.}$       & $^{+0.}_{-0.}$       & $^{+0.018}_{-0.018}$
                  & $-0.426$      & $^{+0.}_{-0.}$       & $^{+0.}_{-0.}$       & $^{+0.019}_{-0.019}$ \\[1mm]
$\vl\gr,\vr\gl$   & $-5.51$       & $^{+0.009}_{-0.010}$ & $^{+0.057}_{-0.057}$ & $^{+0.}_{-0.}$
                  & $-5.48$       & $^{+0.008}_{-0.010}$ & $^{+0.057}_{-0.056}$ & $^{+0.}_{-0.}$
\end{tabular}
\end{center}
\end{footnotesize}
\caption{$\kappa$ factors for the $tb$ and $\bar t b$ processes and their uncertainties, explained in the text.
Errors smaller than $0.005$ are omitted.}
\label{tab:tb}
\end{table}

\section{$\boldsymbol{tW}$ production}
\label{sec:4}

\begin{figure}[htb]
\begin{center}
\begin{tabular}{ccc}
\epsfig{file=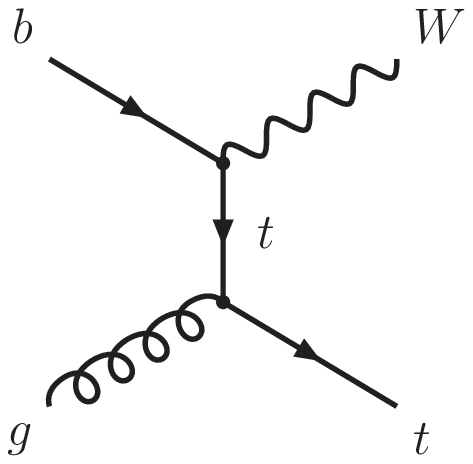,height=3cm,clip=} & \quad \quad &
\epsfig{file=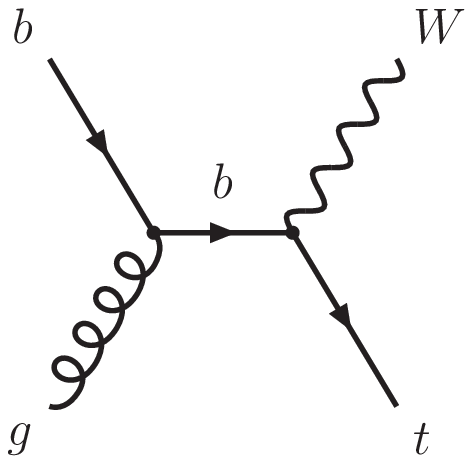,height=3cm,clip=}
\end{tabular}
\caption{Feynman diagrams for single top production in the $gb \to tW^-$ process.}
\label{fig:diag3}
\end{center}
\end{figure}

At lowest order, the $gb \to tW^-$ process is mediated by the two diagrams in Fig.~\ref{fig:diag3}, where the initial $b$ quark comes from splitting $g \to b \bar b$ of a gluon in the proton sea. The charge conjugate process exhibits the same features and has the same cross section, so we will often refer only to $tW^-$ production for brevity. 
Even at this order the $t$ and $b$ quarks entering the $Wtb$ vertex are not on their mass shell in the first and second diagrams, respectively.
Like the $t$-channel process, $tW^-$ production has a NLO correction from
$gg \to tW^-\bar b$. 
\begin{figure}[htb]
\begin{center}
\begin{tabular}{ccccccc}
\epsfig{file=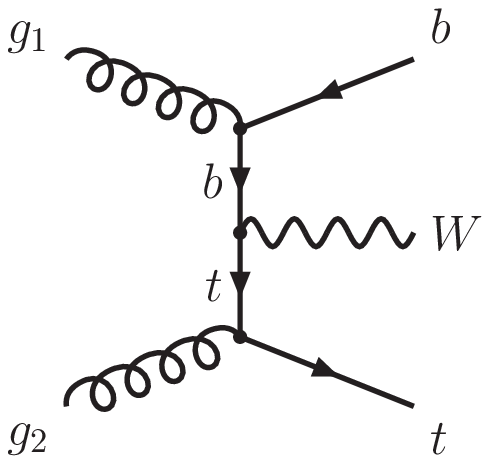,height=2.5cm,clip=} & \quad &
\epsfig{file=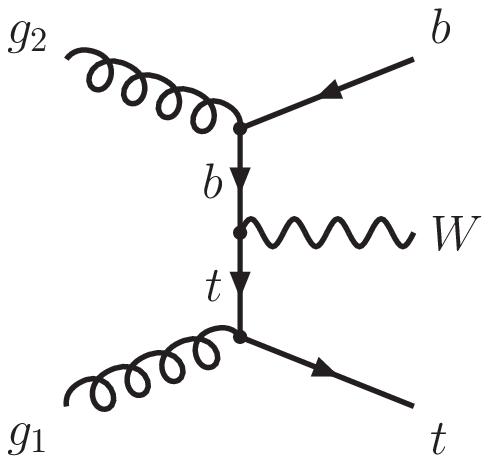,height=2.5cm,clip=} & \quad &
\epsfig{file=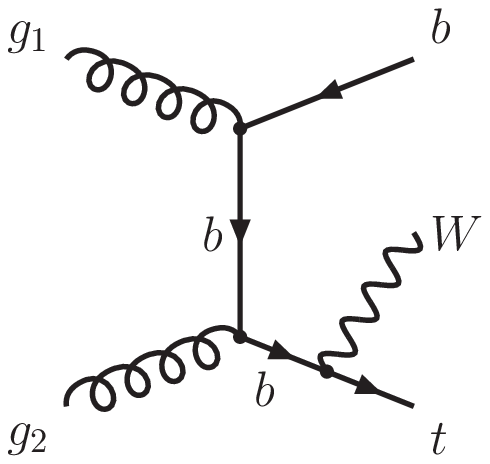,height=2.5cm,clip=} & \quad &
\epsfig{file=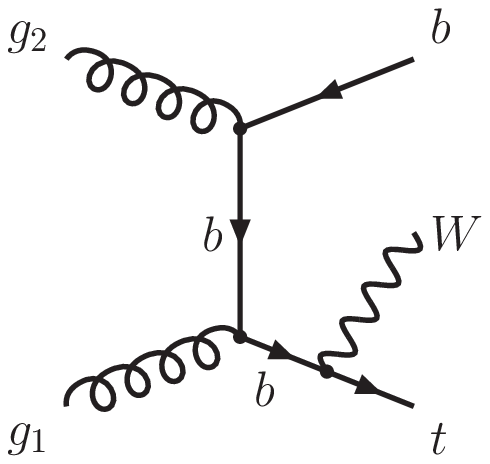,height=2.5cm,clip=} \\[2mm]
\epsfig{file=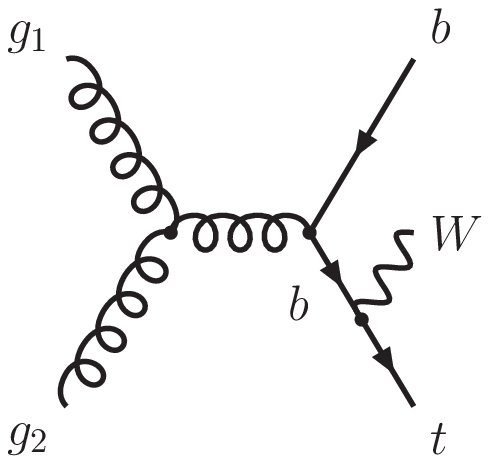,height=2.5cm,clip=} &  &
\epsfig{file=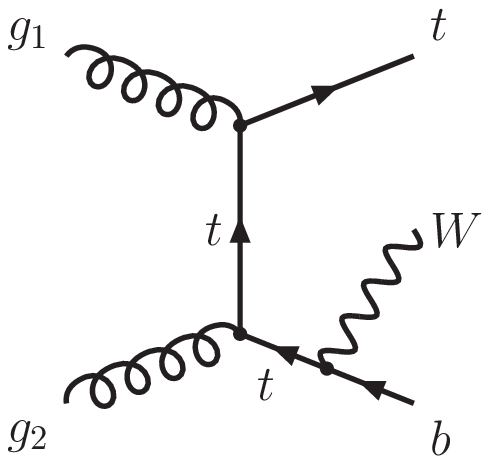,height=2.5cm,clip=} &  &
\epsfig{file=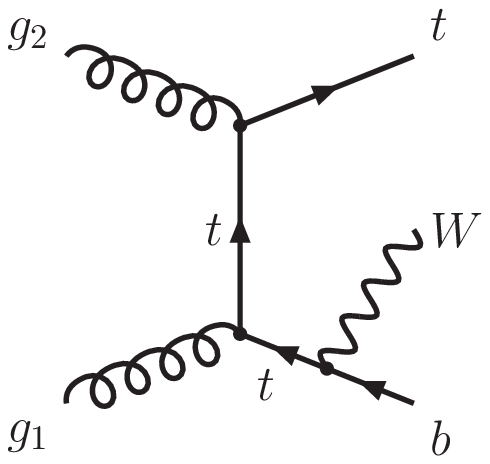,height=2.5cm,clip=} &  &
\epsfig{file=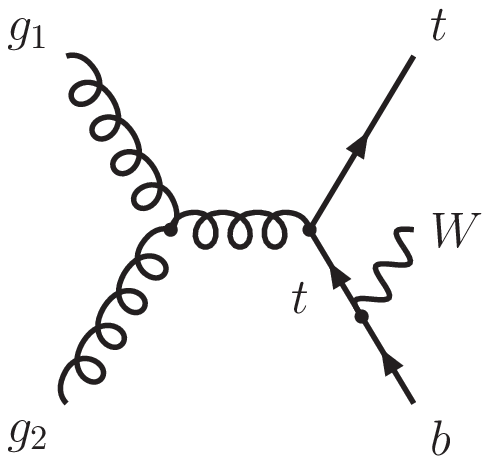,height=2.5cm,clip=}
\end{tabular}
\caption{Feynman diagrams for $gg \to tW^- \bar b$. The two initial gluons are labelled as $g_1$ and $g_2$ for clarity. The last three diagrams correspond to resonant $t \bar t$ production with decay $\bar t \to W^- \bar b$.}
\label{fig:diag4}
\end{center}
\end{figure}
However, an important difference is
that the full gauge invariant set of 8 Feynman diagrams for $gg \to t W^- \bar b$ (see Fig.~\ref{fig:diag4}) also includes on-shell $t \bar t$ production with $\bar t \to W^- \bar b$, whose cross section is about one order of magnitude larger than for $tW^-$ itself. Apart from the displeasing fact of considering $t \bar t$ as a huge ``correction'' to $tW^-$ production, this is not convenient from a practical point of view. The motivation for studying single top production in this paper, and perhaps the main reason to study these processes at all, is to measure $V_{tb}$ and set limits on anomalous $Wtb$ couplings. But the $t \bar t$ cross section is actually independent of the $Wtb$ interaction. Top pairs are produced in a QCD process and, as long as $t \to W b$ dominates the top decay (which is expected, due to the smallness of $V_{td}$ and $V_{ts}$),
the branching ratio is independent as well. It must be pointed out, however, that
if one restricts the invariant mass of the $W^-$ boson and $\bar b$ quark $\mwb$ to select only off-peak contributions, for example requiring $|\mwb-m_t| > 20$ GeV, then the $gg \to t \bar t \to t W^- \bar b$ cross section in this region does depend on the $Wtb$ coupling, {\em e.g.} it scales with $\vl^2$ like in single top production. Still, the total cross section in all phase space remains independent, because the cross section around the peak (whose height is determined by the top width $\Gamma_t$, which depends on $Wtb$ couplings) compensates the variations in the region where the $\bar t$ quark is off-shell.

Since the total $gg \to t \bar t \to t W^- \bar b$ cross section is practically independent of the $Wtb$ interaction,\footnote{Of course, this does not
preclude the fact that the $\bar t \to W^- \bar b$ angular distribution and the $W$ polarisation, as well as the kinematical distributions determined by them, depend on the structure of the $Wtb$ vertex, as it is well known
\cite{Kane:1991bg,Jezabek:1994zv,Jezabek:1994qs,Lampe:1995xb,
Grzadkowski:1999iq,Boos:1999ca,Boos:2001sj,
Grzadkowski:2002gt,delAguila:2002nf,Espriu:2002wx,Grzadkowski:2003tf,
Chen:2005vr,AguilarSaavedra:2006fy,Cao:2006pu,Najafabadi:2006um}.}
if one wants to study the influence of the latter on the total
$t W^- \bar b$ cross section it is natural to consider $gg \to t \bar t \to t W^- \bar b$ as ``background'' and the excess of events over it, which depends on $\vl$ and anomalous couplings, as ``signal''. This approach is convenient from the point of view of the experimental analysis as well. The single top signal is then
\begin{equation}
\sigma_\text{signal} \equiv \sigma_8(gg \to t W^- \bar b) - \sigma_3(gg \to t \bar t
\to t W^- \bar b) \,,
\label{ec:twb1}
\end{equation}
where the first term $\sigma_8$ includes the full set of 8 diagrams in Fig.~\ref{fig:diag4}
and the second term $\sigma_3$ only involves the three resonant $t \bar t$ production diagrams with $\bar t$ decay (the last three diagrams). Clearly,
\begin{equation}
\sigma_\text{signal} = \sigma_5(gg \to t W^- \bar b) + \sigma_\text{int} \,,
\label{ec:twb2}
\end{equation}
so that the $tW^- \bar b$ excess over the $t \bar t$ ``background''
is due to the 5 non-resonant diagrams in Fig.~\ref{fig:diag4} ($\sigma_5$) plus their interference with
$t \bar t$ production (this equation defines $\sigma_\text{int}$).
This interference is not negligible, and amounts to a $-20\%$. Note that this calculation of the single top contribution involving subtraction of $t \bar t$ involves small violations of gauge invariance of order $\Gamma_t/m_t$.

There is a different (gauge invariant) method in the literature \cite{Belyaev:1998dn,Belyaev:2000me} proposed to remove the $t \bar t$ contribution from $t W^- \bar b$, which consists in performing a cut requiring $|\mwb-m_t|$ larger than some quantity, of the order of
$15-20$ GeV. But still with this kinematical cut the $t \bar t$ contribution is comparable to the non-resonant one. In Table~\ref{tab:ttcut} we collect, for various $\mwb$ cuts, the cross sections corresponding to several subsets of diagrams, using CTEQ6L1 PDFs with $Q=m_t+M_W$
\cite{Zhu:2002uj}.\footnote{To our knowledge, there is no study in the literature comparing the leading order (LO) and NLO kinematics for $tW^-$ production, unlike in
$s$- and $t$-channel single top production \cite{Sullivan:2004ie,Boos:2006af}.
Then, we conservatively use LO PDFs for the $tW^-$ and $tW^- \bar b$ processes.}
Notice that
even for $|\mwb-m_t| > 50$ GeV the $t \bar t$ contribution $\sigma_3$ is comparable (a 40\%) to the non-resonant one $\sigma_5$, and their interference is large ($-30\%$). This is not surprising, because the three $t \bar t$ diagrams do not have any particular suppression, and outside the $\mwb$ peak they are expected to have similar size as the non-resonant ones. Thus, $t \bar t$ contributions cannot be fully removed by this method (see also
Ref.~\cite{Kauer:2001sp}). One could still argue that $t \bar t$ production outside the $\mwb$ peak behaves as single top production, because the cross section in this phase space region depends on $\vl$ and anomalous couplings, and might be tempted to include it as single top production. But separating
$t \bar t \to t W^- \bar b$ in two parts, near the peak and off-peak, makes
the former (which is judiciously taken as ``background'') also dependent on $Wtb$ couplings and only complicates the analysis.\footnote{Note however that a cut on $\mwb$ will likely be useful to reduce the $t \bar t$ background and thus the experimental uncertainty associated to the subtraction. Then, in the real analysis a $t \bar t$ subtraction with a $\mwb$ cut may yield the best results. This issue requires a detailed evaluation of systematic uncertainties and is beyond the scope of the present work. }

\begin{table}[ht]
\begin{center}
\begin{tabular}{cccccc}
       & $\sigma_8$ & $\sigma_3$ & $\sigma_8-\sigma_3$ & $\sigma_5$ & $\sigma_\text{int}$ \\
no cut & 439        & 426        & 12.9                & 16.5       & -3.6 \\
15 GeV & 24.7       & 13.6       & 11.1                & 14.5       & -3.5 \\
25 GeV & 18.0       & 8.20       & 9.81                & 13.2       & -3.4 \\
50 GeV & 10.70      & 4.19       & 6.51                & 9.71       & -3.20
\end{tabular}
\end{center}
\caption{Cross sections (in pb) for $g g \to tW^- \bar b$ involving different subsets of
Feynman diagrams: the full set ($\sigma_8$), resonant $t \bar t$ production ($\sigma_3$),
non-resonant diagrams ($\sigma_5$) and interference between $t \bar t$ and non-resonant diagrams ($\sigma_\text{int}$). A kinematical cut on $|m_{Wb}-m_t|$ is applied in all but the first row (see the text). For the charge conjugate process
$gg \to \bar t W^+ b$ the cross sections are equal.}
\label{tab:ttcut}
\end{table}

\begin{figure}[t]
\begin{center}
\begin{tabular}{c}
\epsfig{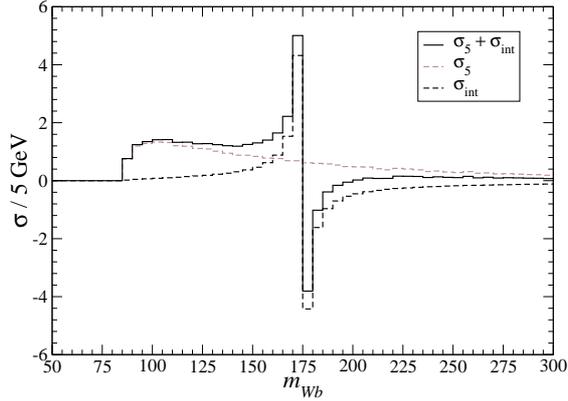}
\end{tabular}
\caption{Kinematical distribution of $\mwb$ for the single top signal in Eq.~(\ref{ec:twb2}) and its two contributions. The cross sections are in pb.}
\label{fig:mwb}
\end{center}
\end{figure}

There is a very interesting feature which can also be observed in the last column of Table~\ref{tab:ttcut}. The net interference between $t \bar t \to t W^- \bar b$ and non-resonant diagrams nearly vanishes around the $\mwb$ peak. Interference terms have the form
\begin{equation}
\text{Re}\, \mathcal{M}_i^* \mathcal{M}_j = 
\text{Re}\, \mathcal{M}_i^*  \mathcal{\tilde M}_j \frac{1}{\mwb^2-m_t^2+i m_t \Gamma_t} \,,
\end{equation}
where $\mathcal{M}_i$, $\mathcal{M}_j$ are the amplitudes of a non-resonant
and resonant diagram, respectively, and $\mathcal{\tilde M}_j$ the latter without the $\bar t$ propagator, which has a momentum $p_{\bar t}^2 = \mwb^2$. Since the two amplitudes do not have relative complex phases (except for the $\bar t$ propagator) the product $\mathcal{M}_{ij} \equiv \mathcal{M}_i^*\mathcal{\tilde M}_j$ is real, and we have
\begin{equation}
\text{Re}\, \mathcal{M}_i^* \mathcal{M}_j = 
\mathcal{M}_{ij} \text{Re}\, \frac{1}{\mwb^2-m_t^2+i m_t \Gamma_t} =
\mathcal{M}_{ij} \frac{\mwb^2-m_t^2}{(\mwb^2-m_t^2)^2 + (m_t \Gamma_t)^2} \,.
\end{equation}
Near the peak at $m_t^2$ the factor $\mathcal{M}_{ij}$ is expected to be approximately constant,
so that integration in $\mwb^2$ within a symmetric interval $[m_t^2-n \Gamma_t,
m_t^2+n \Gamma_t]$ with $n$ not too large
gives a vanishing interference. The cancellation can be nicely observed in Fig.~\ref{fig:mwb}: for $\mwb < m_t$ the interference is positive, while for
$\mwb > m_t$ it is negative, leading to a non-trivial peak-dip structure.
This also implies that $\sigma_\text{int}$ is rather independent of the top width, even if it is changed by a factor of four.\footnote{It must be noted that the effect of anomalous couplings in the top width is much more modest than in single top cross sections due to the different momenta scales involved.}
This feature is crucial for our analysis, because it allows us to safely neglect the dependence of $\Gamma_t$ on anomalous couplings, and justifies expanding
$\sigma_8-\sigma_3$ in terms of $\kappa$ factors and products of $Wtb$ couplings,
as done in Eq.~(\ref{ec:k}). (Non-resonant contributions in $\sigma_5$ are already practically independent of $\Gamma_t$, which is negligible in the denominators.)

\begin{figure}[htb]
\begin{center}
\begin{tabular}{c}
\epsfig{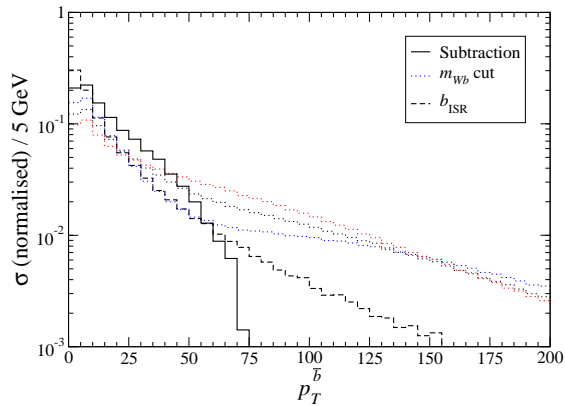}
\end{tabular}
\caption{Normalised transverse momentum distribution for the $\bar b$ quark in the
$tW^-\bar b$ process in several cases: for the hard $tW^-\bar b$ process subtracting the $t \bar t$ contribution (solid line), with a cut on $|\mwb-m_t|$ of 15, 25 or 50 GeV (dotted lines, from top to bottom) and from ISR added to the $tW^-$ process in a parton shower Monte Carlo (dashed line).}
\label{fig:ptbisr2}
\end{center}
\end{figure}

Subtracting the $t \bar t \to t W^- \bar b$ contribution and
considering $\sigma_8-\sigma_3$ as our signal also has some drawbacks. For example, one has to introduce negative weight events, as it is apparent from
Fig.~\ref{fig:mwb}. More importantly, the $\ptb$ distribution does not exhibit a good behaviour, and even becomes negative for $\ptb > 80$ GeV. In Fig.~\ref{fig:ptbisr2} we present this distribution for $t W^- \bar b$ in three cases: (i) when the $t \bar t \to t W^- \bar b$ contribution is subtracted; (ii) when a cut of
15, 25 or 50 GeV is applied on $|\mwb-m_t|$; (iii) when the $\bar b$ quark is generated from ISR added to the $tW^-$ process by {\tt Pythia}. The distribution after $t \bar t$ subtraction is very soft, in contrast with the case when a cut is applied on $\mwb$, indicating that $t \bar t \to t W^- \bar b$ give the $\bar b$ quarks with higher $\ptb$ even away from the peak.

This bad behaviour of the $\ptb$ distribution makes it difficult to perform a matching between $tW^-$ and $tW^-\bar b$ like the one used in section \ref{sec:2} for the $t$-channel process, unless some unrealistic $p_T^\text{cut}$ is chosen. Instead, we remove double counting between $tW^-$ and $tW^-\bar b$ by subtracting the first order logarithmic corrections included in the $2 \to 3$ process \cite{Belyaev:2000me,Tait:1999cf}, calculated by replacing 
in the calculation of the $gb \to tW^-$ cross section
the $b$ quark PDF by the function in Eq.~(\ref{ec:bpdf}).
The subtraction term is rather large \cite{Belyaev:2000me}: using CTEQ6L1 PDFs and $Q=m_t+M_W$ we find $\sigma(gb \to tW^-) = 28.5$ pb, and for the subtraction term
$\sigma_\text{sub} = 21.0$ pb. Notice that
$\sigma(gb \to tW^-) - \sigma_\text{sub} + \sigma_5(gg \to t W^- \bar b) = 24.0$ pb, a 15\% smaller than the first order result for $\sigma(gb \to tW^-)$. The interference with $t \bar t \to tW\bar b$ further decreases the total cross section to 20.4 pb, a 28\% smaller than $\sigma(gb \to tW^-)$.

For $tW^-$ we calculate the $\kappa$ factors as in the previous single top processes, using a squared matrix element calculated analytically with {\tt FORM} (and numerically checked with {\tt HELAS}) in which the different contributions are identified. For $tW^- \bar b$, the
larger number of diagrams (8, giving 36 terms in the squared amplitude) and the number of products of couplings in each make this approach unsuitable. We calculate the $\kappa$ factors numerically by evaluating the squared matrix element iteratively, with different $Wtb$ couplings. Taking one coupling equal to one and the rest zero,
we can obtain the four coefficients of quadratic terms. Then, taking two couplings equal to one and the rest
vanishing allows to extract the six interference terms, subtracting the quadratic contributions previously calculated.
The results are presented in Tables~\ref{tab:tW} and \ref{tab:tWb},
omitting the $\kappa$ coefficients smaller than 0.1. The uncertainties have the same source as in the previous sections: the intervals in the first column correspond to the PDF uncertainty, and the last three columns to the uncertainties from factorisation scale, the top mass and the bottom mass,
respectively.
In these processes both top and antitop production have the same values of $\kappa$, providing an additional cross-check of our calculations.

\begin{table}[ht]
\begin{center}
\begin{tabular}{cclll}
                    & $\kappa$         & $\Delta Q$           & $\Delta m_t$         & $\Delta m_b$         \\[1mm]
$\vr^2$           & $1$           & --                   & --                   & -- \\[1mm]
$\gl^2,\gr^2$     & $3.46-3.57$      & $^{+0.23}_{-0.11}$   & $^{+0.015}_{-0.015}$ & $^{+0.009}_{-0.008}$ \\[1mm]
$\vl \gr,\vr \gl$ & $1$           & --                   & --                   & -- \\[1mm]
\end{tabular}
\end{center}
\caption{Representative $\kappa$ factors for the $t W^-$ and $\bar t W^+$ processes and their uncertainties, explained in the text. Errors smaller than $0.005$ are omitted.}
\label{tab:tW}
\end{table}
\begin{table}[ht]
\begin{center}
\begin{tabular}{cclll}
                    & $\kappa$         & $\Delta Q$           & $\Delta m_t$         & $\Delta m_b$         \\[1mm]
$\vr^2$             & $1$              & --                   & --                   & -- \\[1mm]
$\gl^2,\gr^2$       & $4.51-4.73$      & $^{+0.19}_{-0.04}$   & $^{+0.009}_{-0.027}$ & $^{+0.030}_{-0.}$ \\
$\vl \gr,\vr \gl$   & $1.21-1.23$      & $^{+0.014}_{-0.003}$ & $^{+0.005}_{-0.007}$ & $^{+0.}_{-0.}$ \\
\end{tabular}
\end{center}
\caption{Representative $\kappa$ factors for the $t W^- \bar b$ and $\bar t W^+ b$ processes and their uncertainties, explained in the text.
Errors smaller than $0.005$ are omitted.}
\label{tab:tWb}
\end{table}

\section{Limits on anomalous couplings}
\label{sec:5}

In this section we present estimates of the limits that can be obtained from single top cross section measurements, and also with their combination with 
 top decay observables, which can be measured either in single top or top pair production. We will show that significant bounds on anomalous couplings can be obtained despite the possibility of cancellations and the contamination among the three single top processes. In these estimates we will assume sensitivities for 
single top cross section measurements based on previous literature, and for top decay asymmetries $\Apm$ and helicity ratios $\rhpm$ (see appendix \ref{sec:b}) we will take the experimental uncertainties from $t \bar t$ production. On the other hand, the experimental precision for the measurement of the asymmetry ratio $\rbl$ has not been estimated as yet. Nevertheless, the results are weakly dependent on the sensitivity of this observable, provided that it is better than $\sim 8\%$.
A more complete study will be presented elsewhere, including all experimental systematic uncertainties and the SM background.

\subsection{Limits from single top production}
\label{sec:5.1}

The calculations of sections \ref{sec:2}--\ref{sec:4} have provided us with theoretical expressions for the single top cross sections including anomalous $Wtb$ couplings. In order to link them to real or simulated experimental data, the detection efficiencies 
for each single top process ({\em i.e.} the fraction of events which survive the selection criteria required for the
experimental analysis) must be known. First, because comparing an experimental sample with a theoretical cross section obviously requires to know which fraction of the events produced are actually present in the sample analysed. But also
because different single top processes will contribute to a given final state, and the relative weights determine the dependence on anomalous couplings of the measured cross section.
Efficiencies depend on the final state considered, selection criteria, etc. and must be determined at least with a fast detector simulation. Once these efficiencies are found for each final state of interest, the experimental measurement of the cross section and its uncertainty allow to set limits on anomalous couplings.

This is best explained with an example. Experimentally, the $t$-channel process can be investigated in final states with semileptonic top decay, requiring the presence of a charged lepton, significant missing energy, a $b$-tagged jet (from the top decay), a forward jet with large transverse momentum and no additional jets with large $p_T$ to suppress $t \bar t$ production (see for example Ref.~\cite{Lucotte:977595}). On the other hand, the theoretical calculation of this process is conveniently separated into $tj$ and $t\bar b j$ production. Their respective efficiencies $\varepsilon_{tj}$ and
$\varepsilon_{t \bar bj}$ can be computed with a Monte Carlo simulation, including detector effects.
Then, the predicted number of single top events is
\begin{eqnarray}
N = N_{tj} + N_{t\bar bj} & = &
 \varepsilon_{tj} \, \sigma(tj)_\text{SM} \left( \vl^2 + \kappa_{tj}^{\vr} \, \vr^2 + \kappa_{tj}^{\vl \vr} \, \vl \vr + \dots
 \right) \notag \\
& & + \varepsilon_{t b j} \, \sigma(t \bar bj)_\text{SM} \left( \vl^2 + \kappa_{t b j}^{\vr} \,
 \vr^2 + \kappa_{t bj}^{\vl \vr} \, \vl \vr + \dots \right) \,, \notag \\
& & + \varepsilon_{t b} \, \sigma(t \bar b)_\text{SM} \left( \vl^2 +
 \vr^2 + \kappa_{t b}^{\vl \vr} \, \vl \vr + \dots \right) \,, \notag \\
& & + \varepsilon_{t W} \, \sigma(t W^-)_\text{SM} \left( \vl^2 +
 \vr^2 + \kappa_{t W}^{\vl \vr} \, \vl \vr + \dots \right) \,, \notag \\
& & + \varepsilon_{t W b} \, \sigma(t W^- \bar b)_\text{SM} \left( \vl^2 +
 \vr^2 + \kappa_{t W b}^{\vl \vr} \, \vl \vr + \dots \right) \,,
\label{ec:Neff}
\end{eqnarray}
where we have introduced additional subscripts on the $\kappa$ factors to distinguish the different processes. The last three terms represent the ``contamination'' from the other single top processes, which can be reduced ({\em i.e.} making $\varepsilon_{t b},\varepsilon_{t W},\varepsilon_{t W b} \ll \varepsilon_{t j},\varepsilon_{t b j}$)
with suitable selection criteria.
For antitop production the expression is the same, but substituting the numerical values for those corresponding
to the charge conjugate processes. It is apparent that the functional dependence on $\vl$ and anomalous couplings of the number of events observed is determined not only by the $\kappa$ factors but also by
the SM cross sections and the efficiencies. This approach is different from the usual single top analyses: here all single top processes are considered as signals, with different efficiencies and dependence on anomalous couplings. Then, three final states can be selected trying to isolate each of the processes ($t$-channel, $t \bar b$ and $tW^-/tW^- \bar b$) as much as possible. Nevertheless, contributions from all processes will always be present, especially from the $t$-channel process which has a much larger cross section.

A second issue to be considered is the dependence of efficiencies on $Wtb$ couplings. Anomalous couplings change the angular and energy distributions both in the production and the decay of the top quark, and then they affect the efficiencies. A complete scan over the 3-dimensional space of anomalous couplings to parameterise the efficiencies is involved, but a simpler
approach can also be followed. First, the SM efficiencies can be used to obtain bounds on anomalous couplings. With these bounds the efficiencies can be reevaluated if a positive signal beyond the SM is found and, in any case,
their uncertainties due to anomalous couplings must be estimated and included in the final results if they are found relevant. A similar procedure is foreseen for the study of anomalous couplings in top decays \cite{Hubaut:2005er}.

For these evaluations we have developed the Monte Carlo generator {\tt Protos} for the different single top production processes considered, including the decay of the top quark and $W$ boson(s) with finite width and spin effects. Matrix elements
for
\begin{align}
& bq \to t q' \to W^+ b q' \to \ell^+ \nu b q' \,, \notag \\
& gq \to t q' \bar b \to W^+ b q' \bar b \to \ell^+ \nu b q' \bar b \,, \notag \\
& q \bar q' \to t \bar b \to W^+ b \bar b \to \ell^+ \nu b \bar b \,, \notag \\
& gb \to t W^- \to W^+ b W^- \to \ell^+ \nu b \bar q  q' \,, \notag \\
& gg \to t W^- \bar b \to W^+ b W^- \bar b \to \ell^+ \nu b \bar q  q' \bar b
\end{align}
and their charge conjugate
are calculated with {\tt HELAS}, with $Wtb$ anomalous couplings implemented in the production as well as in the decay of the top quark. The output of the generators provide events with the colour information necessary for hadronisation in order to be interfaced to {\tt Pythia}. These generators are also used to check the effect of the top quark decay in the $\kappa$ factors, which is discussed in appendix \ref{sec:c}.

We calculate the efficiencies $\varepsilon$ using a fast simulation of the ATLAS detector \cite{atlfast}. We restrict ourselves to the electron channel, final states with a muon are alike. Event samples for each the process are generated for final states containing at least one $e^\pm$, corresponding to luminosities
of 30 fb$^{-1}$ for $tj$ and $t \bar b j$, 300 fb$^{-1}$ for $t \bar b$ and
60 fb$^{-1}$ for $tW^-$ and $tW^- \bar b$. For the latter process, this requires simulating about two times more events, so that the sum of events with positive and negative weight corresponds to the cross section times the luminosity.
Results are finally rescaled to 30 fb$^{-1}$. A $K$ factor of 1.54
\cite{Zhu:2002uj,Campbell:2005bb} is introduced in the $tW^-$, $tW^- \bar b$ processes for consistency with the other channels, resulting in
cross section $\sigma(tW^- + tW^- \bar b) = 31.4$ pb. Events are passed through {\tt Pythia} including pile-up, {\tt ATLFAST} and {\tt ATLFASTB}, where a $b$ tagging efficiency of 60\% is chosen. We concentrate ourselves on a final state for each single top production process, requiring in all of them the presence of an isolated electron or positron with $p_T > 25$ GeV and missing energy larger than 25 GeV. The rest of selection criteria,
adopted to isolate each process as far as possible, are respectively:
\begin{enumerate}
\item[(i)] Final state 1 (for the $t$-channel process): a forward jet with pseudorapidity in the range $2.5 < |\eta| < 5$ with $p_T > 50$ GeV; at least one central $b$ jet with
$p_T > 30$ GeV; at most one additional central jet, which cannot have
$p_T > 30$ GeV. The top quark mass is reconstructed from the charged lepton and $b$ quark momenta (if there are more than one $b$ jet we select the one with largest transverse momentum), and the missing energy. The transverse neutrino momentum is assumed to equal the missing energy of the event, and the longitudinal component is found solving a quadratic equation requiring that $(p_e + p_\nu)^2 = M_W^2$.
The solution selected is the one with smaller $p_{\nu z}$. Once that the top quark momentum and its invariant mass are reconstructed, the latter is required to be between 150 GeV and 225 GeV.
\item[(ii)] Final state 2 (for $t \bar b$ production): we require that neither forward jets nor light central jets are present with transverse momentum larger than 15 GeV. Final states must have two $b$-tagged jets with $p_T > 30$ GeV.
\item[(iii)] Final state 3 (for $tW^-$ and $tW^- \bar b$ production): the signal is enhanced requiring one (and only one) $b$ jet with $p_T > 30$ GeV and at least two light jets with $p_T > 50$ GeV.
The invariant mass of the $W$ decaying hadronically is reconstructed from the two light jets with the largest transverse momentum, and it must be between 60 and 110 GeV.
\end{enumerate}
The number of events for the five production processes considered are listed in Table~\ref{tab:ev}, before and after the selection criteria for the final states
1 ($tj$), 2 ($t \bar b$) and 3 ($tjj$), respectively.
Comparing with Ref.~\cite{Lucotte:977595}, from where we take the experimental (statistical plus systematic) uncertainties expected, we observe that these selection criteria reproduce the features most important for our analysis: (i) the $t$-channel process, with a cross section much larger, can be cleanly separated from the other two processes; (ii) the $t \bar b$ process has a large contamination from $t$-channel single top production, which is about two times larger even after
selection criteria are applied; (iii) $tW^- + tW^-\bar b$ production can also be separated from the other processes, and is about 4 times larger after selection.
\begin{table}[ht]
\begin{center}
\begin{tabular}{ccccc}
& No sel & sel 1 & sel 2 & sel 3 \\[1mm]
$tj$          & 391200 & 7914  & 398   & 570 \\
$t \bar b j$  & 430800 & 6231  & 1411  & 1265 \\
$t \bar b$    & 35500  & 271.2 & 957.7 & 120.0 \\
$tW^- $       & 145200  & 250.5 & 106.0 & 3455 \\
$tW^- \bar b$ & 249800 & 392.0 & 111.5  & 4869.5 \\
\end{tabular}
\end{center}
\caption{Number of events for the different single top production processes in
final states containing at least one $e^\pm$ (first column) and after the selection criteria corresponding to the three final states studied. The luminosity assumed is 30 fb$^{-1}$.}
\label{tab:ev}
\end{table}
Exact agreement with Ref.~\cite{Lucotte:977595} is not expected, because the signal modeling is different (only $t \bar b j$ is considered in the $t$-channel process and the $t W^- \bar b$ correction is not added to $tW^-$ production) and pile-up was not included in those simulations. For these reasons we have not applied the same selection criteria but have slightly adapted them. Optimisation of the
cuts is of course possible, as well as the use of a likelihood analysis to further suppress the SM background, mainly constituted by top pair production. This is beyond the scope of the present discussion.

The expressions of the cross sections for arbitrary $Wtb$ couplings have been implemented in {\tt TopFit} \cite{AguilarSaavedra:2006fy}, including their theoretical uncertainties. The efficiencies for the different final states are calculated from Table~\ref{tab:ev}, and it is assumed that the number of measured events equals the SM expectation. The experimental errors assumed for cross section measurements in the three final states studied are \cite{Lucotte:977595}:
\begin{align}
&\mbox{Final state 1 ($tj$):} && 1.0\% \,\text{(stat)} \oplus 11\% \,\text{(exp)}
\oplus 6\% \,\text{(bkg)} \oplus 5\% \,\text{(lum)} \,, \notag \\
&\mbox{Final state 2 ($t \bar b$):} && 12\% \,\text{(stat)} \oplus 12\% \,\text{(exp)} \oplus 11\% \,\text{(bkg)} \oplus 5\% \,\text{(lum)} \,, \notag \\
&\mbox{Final state 3 ($tjj$):} && 1.5\% \,\text{(stat)} \oplus 11\% \,\text{(exp)} \oplus 9.1\% \,\text{(bkg)} \oplus 5\% \,\text{(lum)} \,,
\label{ec:errsin}
\end{align}
where in each final state the first term corresponds to the statistical error (including the background); the second is the experimental uncertainty from jet energy scaling, $b$ tagging, etc.; the third one arises from background normalisation, and the last one from the luminosity determination.
Errors are summed in quadrature, and amount to a 13.5\%, 20.8\% and 15.2\%,
respectively. If, in the real experiment, these errors can be decreased, {\em e.g.} with a jet energy calibration better than expected or with a different reconstruction method, then the results will be correspondingly better.
The experimental uncertainty for the ratio $R(\bar t/t)$ in the final state 1 ($tj$) has been not estimated yet. We assume a 2\% statistical error, consistent with the one in the first of Eqs.~(\ref{ec:errsin}).
The luminosity uncertainty cancels in the ratio, which is 
also less sensitive to the background normalisation because the largest background is $t \bar t$ production, which contributes equally to $t$ and
$\bar t$ final states.
Experimental errors from jet energy scaling, $b$ tagging, etc. will likely affect the ratio to a lesser extent. For these uncertainties we tentatively assume a global value of 3\%, to be summed in quadrature to the statistical error.
Correlations between experimental systematic uncertainties in the different final states must be evaluated with a dedicated analysis, and have not been included in the limits.
The correlation between $\sigma(t+\bar t)$ and $R(\bar t/t)$ vanishes if the numbers of events observed $N_t$, $N_{\bar t}$ follow Gaussian statistics with standard deviations $\sqrt N_t$, $\sqrt N_{\bar t}$.
The statistical derivation of limits on anomalous couplings from observables is done in {\tt TopFit} with the acceptance-rejection method, as explained in Refs.~\cite{AguilarSaavedra:2006fy,AguilarSaavedra:2007rs}, and the limits presented
here correspond to a confidence level of 68.3\%.

Before discussing the combined limits on the four $Wtb$ couplings it is illustrative to consider examples in which some parameters are restricted to their SM values.
\begin{figure}[t]
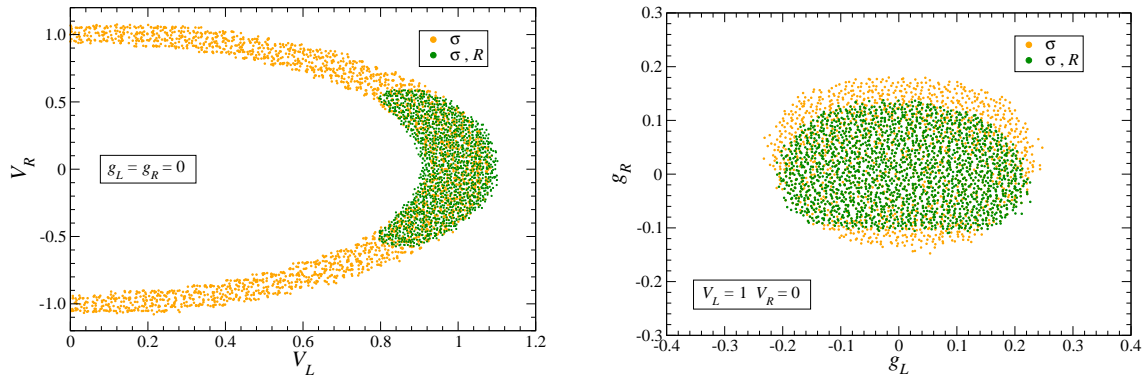

\begin{center}
\begin{tabular}{ccc}
\epsfig{file=Figs/ex1-vlvr.eps,height=4.9cm,clip=} & \quad &
\epsfig{file=Figs/ex1-glgr.eps,height=4.9cm,clip=}
\end{tabular}
\end{center}
\caption{Estimated two-dimensional limits (with 68.3\% CL) on $(\vl,\vr)$
and $(\gl,\gr)$, obtained
from measurement of single top cross sections, with and without the
ratio $R(\bar t/t)$ for the $tj$ final state.}
\label{fig:ex1}
\end{figure}
In Fig.~\ref{fig:ex1} we display two-dimensional limits in the case that all couplings except the pairs $(\vl,\vr)$, $(\gl,\gr)$, respectively, take their SM values. We distinguish the limits obtained only from cross sections (in the three channels) and when the ratio $R(\bar t/t)$ is also included. The most interesting features of these results are:
\begin{itemize}
\item[(i)] Limits on $(\vl,\vr)$ reveal that, as anticipated in section \ref{sec:2}, the measurement of single top cross sections alone cannot discriminate $\vl$ against $\vr$, and the ratio $R(\bar t/t)$ must be included as well.
This ratio is equivalent to the asymmetry proposed in Ref.~\cite{Boos:1999dd}.
\item[(ii)] The inclusion of $R(\bar t/t)$ gives a moderate improvement in
the limits on $(\gl,\gr)$.
\end{itemize}

\begin{figure}[ht]
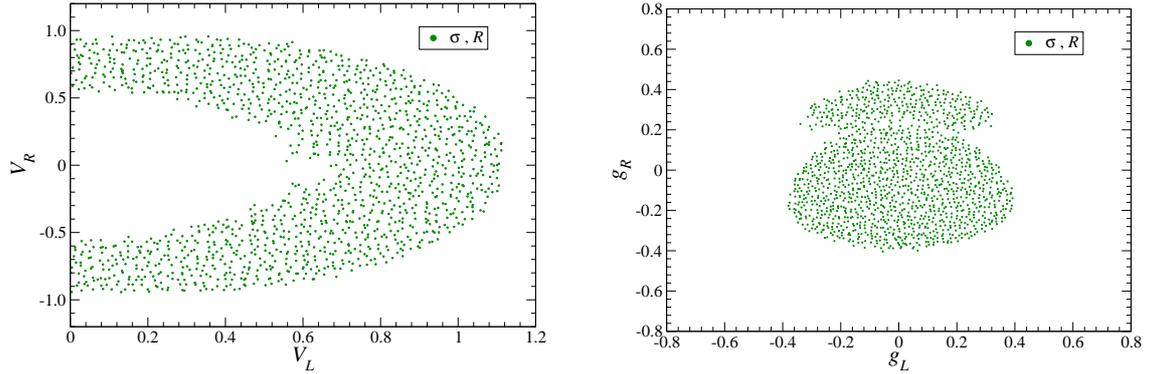

\begin{center}
\begin{tabular}{ccc}
\epsfig{file=Figs/fit-XR-vlvr.eps,height=4.9cm,clip=} & \quad &
\epsfig{file=Figs/fit-XR-glgr.eps,height=4.9cm,clip=}
\end{tabular}
\end{center}
\caption{Combined limits on $Wtb$ couplings from single top cross section measurements, including  $R(\bar t/t)$. The two graphs correspond to different projections of the 4-dimensional allowed region (with 68.3\% CL).}
\label{fig:fitXR}
\end{figure}

The combined limits obtained leaving the four parameters arbitrary are presented in Fig.~\ref{fig:fitXR}.  Bounds on $\vl$ and $\vr$ are rather loose even including $R(\bar t/t)$, because of cancellations with terms involving $\gl$ and especially $\gr$. Limits on $\gl$ and $\gr$
 are also weaker than the corresponding ones
in Fig.~\ref{fig:ex1}, up to a factor of two.

Some comments regarding these limits are in order. As it can be clearly seen from the left plot in Fig.~\ref{fig:fitXR}, single top cross section measurements {\em will not}
provide a measurement of $\vl$ by themselves, unless additional assumptions
on the rest of couplings are made. Notice that even setting $\vr=0$, as it is often done in the literature, gives a large interval $[0.5,1.2]$ for $\vl$ at 68.3\% CL. 
The large allowed range for $\vl$ is caused by the (partial) cancellations
among the contributions to the cross sections involving $\vl$, $\vr$, $\gl$ and $\gr$, and the limited experimental sensitivity. In this way, sets of these parameters very different from the SM values $\vl=1$, $\vr=\gl=\gr=0$ give approximate agreement with the SM predictions for cross sections in the three channels. The weakness of the limits obtained emphasises the importance of a combination with top decay observables, which will be carried out in the next subsection.

\subsection{Combination with top decay observables}
\label{sec:5.2}

Angular distributions and asymmetries in top decays
are in many ways complementary to single top cross sections, and the combination of both provides much stronger limits on $Wtb$ couplings.
While the former cannot fix the value of $\vl$, they are much more sensitive to $\vr$, $\gl$ and $\gr$ than the latter.
We will not address here the interplay among single top production and top decay distributions in detail. This discussion will be presented elsewhere, when all the details regarding the experimental sensitivity for the observables involved will be studied. Still, it is very interesting to know the possible result of a combination, not only because of the better limits obtained, but also to test whether top decay asymmetries help improve the limits on $Wtb$ couplings so that the efficiency variations in single top production due to anomalous couplings are small.

In this combination we assume for the observables $\Apm$ and $\rhpm$
the experimental sensitivities of
top pair production. These observables are redundant but they can be combined as long as their correlation matrix is not singular \cite{AguilarSaavedra:2007rs}.
The combination of $\Apm$, $\rhpm$ measurements in top pair and single top production will certainly improve the sensitivities with respect to these values and give tighter constraints on anomalous couplings.
For the asymmetry ratio $\rbl$ we tentatively assume an experimental error of 2\%, uncorrelated with the former.
Results are not strongly dependent on the $\rbl$ precision, as it will be demonstrated at the end of this subsection.
We do not include $\rnl$ in the fits, because it gives no improvement over $\rbl$ alone unless their experimental uncertainties are similar. This is unlikely, due to the difficulty in reconstructing the momentum of the missing neutrino. For the combination we do not include $R(\bar t/t)$ either. This observable is superseeded by
$\Apm$ and $\rhpm$, and it would add some extra information to the fits only if its experimental precision was much better than the one estimated in the previous subsection.

\begin{figure}[ht]
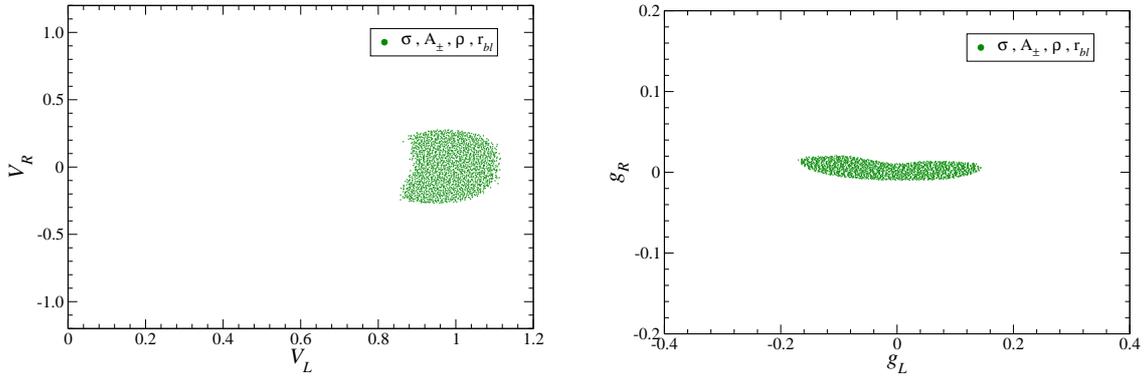

\begin{center}
\begin{tabular}{ccc}
\epsfig{file=Figs/fit-AX-vlvr.eps,height=4.9cm,clip=} & \quad &
\epsfig{file=Figs/fit-AX-glgr.eps,height=4.9cm,clip=}
\end{tabular}
\end{center}
\caption{Combined limits on $Wtb$ couplings from single top cross section measurements (excluding  $R(\bar t/t)$) and top decay observables $\Apm$, $\rhpm$, $\rbl$. The two graphs correspond to different projections of the
4-dimensional allowed region (with 68.3\% CL).}
\label{fig:fitall}
\end{figure}

The combined limits obtained from the expected measurements of single top cross sections, $\Apm$, $\rhpm$ and $\rbl$ are presented in Fig.~\ref{fig:fitall}.
These limits are far better than the ones obtained either with single top cross sections or top decay observables alone, and clearly show the benefit of the combination among them. In particular,
\begin{itemize}
\item[(i)] $\vl$ is bounded with a relatively good precision, $0.85 \leq \vl \leq 1.11$ at one sigma, only a factor of 1.5 worse than the limit $0.92 \leq \vl \leq 1.10$
that can be obtained from single top cross section measurements under the assumption that all anomalous couplings vanish.
\item[(ii)] The constraints on $\vr$ and $\gl$ are moderately strong, due to a fine-tuned cancellation between them in $\Apm$ and $\rhpm$. This cancellation is decreased by the measurement of $\rbl$ with a precision of $\sim 8\%$ or better, as argued below.
\item[(iii)] Limits on $\gr$ are very stringent, $-0.012 \leq \gr \leq 0.024$, even as good as the ones which have been previously obtained from top decay observables for $\vl=1$ and assuming no cancellation between $\vr$ and $\gl$ \cite{AguilarSaavedra:2007rs}.
\end{itemize}

\begin{figure}[ht]
\begin{center}
\epsfig{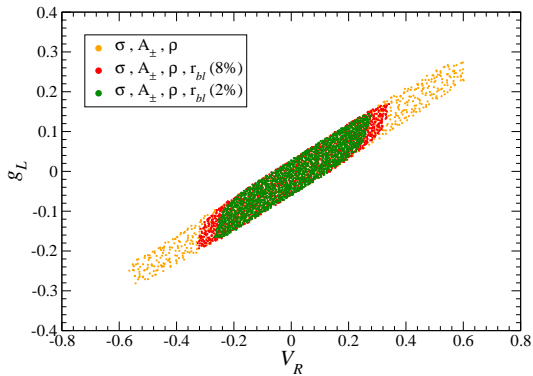}
\end{center}
\caption{Projection on the $(\vr,\gl)$ plane of the combined limits on $Wtb$ couplings from single top cross section measurements and top decay observables $\Apm$, $\rhpm$, without $\rbl$ and with precisions of 8\%, 2\%.}
\label{fig:vrgl}
\end{figure}

We finally point out the important role of the asymmetry ratio $\rbl$, introduced in this paper, in order to improve limits on $Wtb$ couplings. In Fig.~\ref{fig:vrgl} we display the projection on
the $(\vr,\gl)$ plane of the 4-dimensional combined limits, distinguishing the cases when $\rbl$ is not measured or it is measured with precisions of 8\% and 2\%. In the first case we can clearly observe the fine-tuned cancellation between $\vr$ and $\gl$ present in the observables $\Apm$
and $\rhpm$. This cancellation was already pointed out in Ref.~\cite{AguilarSaavedra:2007rs}. Single top cross section measurements do not significantly reduce it, but when the $\rbl$ measurement is added to the fits the cancellation is greatly decreased, even if the precision in the measurement is not very high. Thus, the limits presented in Fig.~\ref{fig:fitall} will likely be maintained to a large extent even if the 2\% goal for $\rbl$ cannot be achieved.

\subsection{Efficiency variation with anomalous couplings}
\label{sec:5.3}

With the limits in Figs.~\ref{fig:fitall} and \ref{fig:vrgl} the efficiency
variation in single top production can be evaluated, using some benchmark
points. We select the following parameter sets, with anomalous couplings in the
boundary of the 68.3\% CL regions:
\begin{align}
& \mbox{Set A} && \vl = 1 \,,\quad \vr=0.3 \,,\quad \gl=0.15 \,, \notag \\
& \mbox{Set B} && \vl = 1 \,,\quad \gr=0.024  \,, 
\end{align}
with the rest of couplings not explicitly written taken to zero. Events are
generated for the $tj$, $t \bar b j$, $t \bar b$, $tW^-$ and $tW^- \bar b$
processes (and their charge conjugate) with each set of couplings, and
simulated as before. The selection criteria (1--3) defined to isolate $t$-channel, $t \bar b$ and $tW^-$ production, respectively, are applied. The
resulting efficiencies are gathered in Table~\ref{tab:eff}, for the SM and the
two sets of anomalous couplings. The quoted errors correspond to the Monte
Carlo statistical error.
\begin{table}[ht]
\begin{center}
\begin{tabular}{cccc}
& \multicolumn{3}{c}{Final state 1 ($tj$)} \\[1mm]
& SM & Set A & Set B  \\[1mm]
$tj$          & $2.02 \pm 0.02$   & $2.08 \pm 0.02$    & $2.07 \pm 0.02$   \\
$t\bar b j$   & $1.44 \pm 0.02$   & $1.40 \pm 0.02$    & $1.46 \pm 0.02$   \\
$t \bar b$    & $0.76 \pm 0.01$   & $0.73 \pm 0.01$    & $0.74 \pm 0.01$   \\
$tW^-$        & $0.173 \pm 0.008$ & $0.161 \pm 0.007$  & $0.173 \pm 0.008$ \\
$tW^- \bar b$ & $0.146 \pm 0.004$ & $0.150 \pm 0.004$  & $0.151 \pm 0.004$ \\[5mm]
\hline \\
& \multicolumn{3}{c}{Final state 2 ($t \bar b$)} \\[1mm]
& SM & Set A & Set B \\[1mm]
$tj$          & $0.102 \pm 0.005$ & $0.103 \pm 0.005$  & $0.104 \pm 0.005$ \\
$t \bar b j$  & $0.328 \pm 0.009$ & $0.315 \pm 0.009$  & $0.328 \pm 0.009$ \\
$t \bar b$    & $2.70 \pm 0.03$   & $2.75 \pm 0.03$    & $2.70 \pm 0.03$   \\
$tW^-$        & $0.073 \pm 0.005$ & $0.082 \pm 0.005$  & $0.065 \pm 0.005$ \\
$tW^- \bar b$ & $0.071 \pm 0.003$ & $0.071 \pm 0.003$  & $0.071 \pm 0.003$ \\[5mm]
\hline \\
& \multicolumn{3}{c}{Final state 3 ($tjj$)} \\[1mm]
& SM & Set A & Set B \\[1mm]
$tj$          & $0.146 \pm 0.006$ & $0.158 \pm 0.006$  & $0.132 \pm 0.006$ \\
$t \bar b j$  & $0.294 \pm 0.008$ & $0.313 \pm 0.009$  & $0.292 \pm 0.008$ \\
$t \bar b$    & $0.33 \pm 0.10$   & $0.38 \pm 0.10$    & $0.32 \pm 0.10$   \\
$tW^- $       & $2.38 \pm 0.03$   & $2.47 \pm 0.03$    & $2.39 \pm 0.03$   \\
$tW^- \bar b$ & $2.37 \pm 0.02$   & $2.38 \pm 0.02$    & $2.36 \pm 0.02$   \\
\end{tabular}
\end{center}
\caption{Efficiencies ($\times 100$) for the different single top production processes in the three final states studied (after selection criteria), within the SM and for two sets of anomalous couplings.}
\label{tab:eff}
\end{table}
For sets A and B efficiency variations are rather mild
for the relevant contributions. (Processes contributing marginally have larger
variations partly due to statistics, but these variations are irrelevant for
the limits finally obtained.) In the $tj$ final state, $t$-channel processes
have efficiency variations smaller than 3\%, well below the 13.5\% experimental
uncertainty. In the $t\bar b$ final state the $t\bar b$ and $t\bar b j$
processes have a variation up to 4.1\%, which is also smaller than the
experimental error of 20.8\%, and in the $tjj$ final state the $tW^-$ and
$tW^- \bar b$ have a maximum variation of 3.8\%, to be compared with the 15.2\%
experimental error.
Hence, considering the efficiency as the SM one is a good first approximation,
although the efficiency variation can be taken into account in a more detailed
analysis.

\section{Conclusions}
\label{sec:6}

Models beyond the SM allow for new physics effects in the $Wtb$ vertex, either deviations from unity in the CKM matrix element $V_{tb}$ \cite{AguilarSaavedra:2002kr,delAguila:2001pu} or $Wtb$ anomalous couplings, generated radiatively \cite{Cao:2003yk,Wang:2005ra} or from new physics at a higher scale.
The main purpose of this paper has been to obtain expressions for single top cross sections at LHC involving arbitrary $Wtb$ couplings. For definiteness we have worked within the framework of gauge invariant effective operators, which allows to relate the $Wtb$ and $gWtb$ couplings from gauge invariance. SM extensions at the electroweak scale can also give radiative corrections to both triple and quartic vertices, and in principle the relation between $Wtb$ and $gWtb$ may not be exactly the one predicted by the
$\mathrm{SU}(2)_L \times \mathrm{U}(1)_Y$ symmetry, which is broken at low energies.
However, in this case the analysis becomes model-dependent and so the framework of gauge invariant effective operators remains simpler.

We have explicitly shown that the only relevant $Wtb$ couplings for single top production are the usual $\gamma^\mu$ and $\sigma^{\mu \nu} q_\nu$ terms, despite the fact that in some of the production processes the top and/or bottom quarks involved in the $Wtb$ interaction are far from their mass shell. For this,
we have introduced a general parameterisation of the $Wtb$ vertex for off-shell top and bottom quarks, in terms of the on-shell Lagrangian (with $\gamma^\mu$ and $\sigma^{\mu \nu}q_\nu $ terms) plus two ``off-shell'' operators $\mathcal{O}_{1,2}$ which vanish when the top and bottom quarks are on their mass shell. Assuming that these couplings arise from gauge invariant operators,
$\mathcal{O}_{1,2}$ have quartic $gWtb$ vertices $\mathcal{O}_{1,2}^{(4)}$ associated by gauge symmetry, which also contribute to single top production. We have seen that the combined contributions involving $\mathcal{O}_{1,2}$ and $\mathcal{O}_{1,2}^{(4)}$
identically cancel, making the single top cross sections independent of the off-shell couplings. This cancellation reduces the number of relevant parameters in the $Wtb$ vertex from 8 to 4 and thus simplifies setting limits from single top cross sections. Furthermore, it also implies that Gordon identities can actually be used on gauge invariant new physics contributions to the $Wtb$ vertex, rewritting $k^\mu$ and $\sigma^{\mu \nu} k_\mu$ terms even for off-shell $t$ and $b$.

We have provided expressions for single top cross sections in terms of $Wtb$ anomalous couplings for (i) the $t$-channel process, including $tj$ and $t \bar b j$, for which a matching has been performed to avoid double counting; (ii) $s$-channel $t \bar b$ production; (iii) $tW^-$ associated production including the $tW^- \bar b$ correction, from which the resonant $t \bar t$ contribution is subtracted at the cross section level. These cross sections have been written as a sum of products of $Wtb$ couplings times numerical coefficients, 
\begin{eqnarray}
\sigma & = & \sigma_\text{SM} \left( \vl^2 + \kappa^{\vr} \, \vr^2 
+ \kappa^{\vl \vr}\, \vl \vr 
+ \kappa^{\gl} \, \gl^2 + \kappa^{\gr}\, \gr^2 + \dots \right)  \,,
\label{ec:k2}
\end{eqnarray}
with the couplings defined in Eqs.~(\ref{ec:1}) and the dots standing for further interference terms.
These expressions can be used to fit the LHC measurement of single top cross sections and simultaneously obtain a measurement of $V_{tb}$ and bounds on anomalous couplings. We have shown how this could be done, taking into account the theoretical uncertainty in single top cross sections and $\kappa$ factors, as well as the expected experimental uncertainty in cross sections \cite{Lucotte:977595}. A difference with respect to usual single top analyses is that here all single top processes are considered as signals, with different efficiencies and also with a different dependence on anomalous couplings.
Indeed, this seems the most reasonable approach in an analysis aiming to measure $V_{tb}$ and anomalous couplings, since the cross section of all single top processes depend on them. Backgrounds are then constituted by processes with cross sections independent of the $Wtb$ interaction, as for example
$t \bar t$ and $W/Z$ production plus jets.

The limits obtained using only single top cross sections are not very strong
because of the experimental uncertainties assumed, of 13\%, 21\% and 15\%
for the $t$-channel, $t \bar b$ and $tW^-$ processes, respectively (which may be a little conservative). Moreover, cancellations among the different couplings, which have not been considered in previous literature, prevent us from obtaining precise bounds using only cross section measurements. Nevertheless, limits
can be greatly improved with the combination with top decay observables, such as the angular asymmetries $\Apm$, helicity ratios $\rhpm$ \cite{AguilarSaavedra:2006fy} and a spin asymmetry ratio $\rbl$ introduced here. We have performed the combination using the expected precision for $\Apm$ and $\rhpm$ in top pair production \cite{AguilarSaavedra:2007rs} and an estimate of the sensitivity in the $\rbl$ measurement. It must be remarked that the limits obtained make no assumption on the $Wtb$ couplings (except that they are CP-conserving). Due to the non-linearity of the equations used and the limited precision of the experimental observables, determining the $Wtb$ couplings without ambiguities is more involved than merely counting parameters and finding an equal number of independent observables.
Actually, it is non-trivial to find stringent limits on $Wtb$ couplings avoiding the fine-tuned cancellations that can occur, mainly between $\vr$ and $\gl$. The asymmetry ratio $\rbl$ (or any equivalent observable) plays a key role in improving the limits, as demonstrated in Fig.~\ref{fig:vrgl}.

The results obtained from the global fit are very promising. Precise bounds on several couplings can be achieved, as well as a measurement of $V_{tb}$ which is only a factor of 1.5 less precise than the (model-dependent) one obtained setting all anomalous couplings to zero. These results have been shown to depend weakly on the sensitivity to $\rbl$ if it is better than $\sim 8\%$, which is likely to happen given the sensitivities of other top decay observables. This combination deserves a detailed investigation when the expected experimental precision of all the observables involved is known in one or more top production and decay processes, and it will be presented elsewhere.

\vspace{1cm}
\noindent
{\Large \bf Acknowledgements}
\vspace{0.3cm}

\noindent
I thank N. Castro, A. Onofre, M. P\'erez-Victoria, R. Pittau and especially F. del Aguila and R. Santos for several useful discussions. This work has been supported by a MEC Ram\'on y Cajal contract, MEC project FPA2006-05294 and
Junta de Andaluc{\'\i}a projects FQM 101 and FQM 437.

\appendix

\section{Cancellation of off-shell operator contributions}
\label{sec:a}

In this appendix we explicitly prove the cancellation among off-shell
triple and quartic contributions to the $g b \to tW^-$ and $gg \to tW^- \bar b$
amplitudes. We first rewrite $\mathcal{O}_{1,2}$ in Eqs.~(\ref{ec:3}) expanding the $\sigma^{\mu \nu}$ terms in both operators, obtaining
\begin{eqnarray}
\mathcal{O}_1 & = & -\frac{g}{\sqrt 2 M_W} \, \bar b  \left[- (\glp P_L + \grp P_R) \gamma^\mu (p_t\!\!\!\!\!\!\!\not\,\,\,\,
- m_t) \right. \nonumber \\
& & \left. + (p_b\!\!\!\!\!\!\!\not\,\,\,\, - m_b) \gamma^\mu (
\glp P_L + \grp P_R) \right] t \; W_\mu^- + \mathrm{H.c.} \,, \nonumber \\
\mathcal{O}_2 & = & -\frac{g}{\sqrt 2 M_W} \, \bar b  \left[ (\hl P_L + \hr P_R) \gamma^\mu (p_t\!\!\!\!\!\!\!\not\,\,\,\,
- m_t) \right. \nonumber \\
& & \left. + (p_b\!\!\!\!\!\!\!\not\,\,\,\, - m_b) \gamma^\mu (
\hl P_L + \hr P_R) \right] t \; W_\mu^- + \mathrm{H.c.} \,,
\label{ec:Oalt}
\end{eqnarray}
where $p_t$ and $p_b$ are the momenta of the quarks involved in the vertex, following the fermion flow.
Note that in deriving Eqs.~(\ref{ec:Oalt}) from the definition in Eqs.~(\ref{ec:3}) we have not used Gordon identities, so that the two sets of equations are completely equivalent for $t$, $b$ off-shell.
These alternative expressions for $\mathcal{O}_{1,2}$
are extremely useful to prove the cancellation among diagrams.
Incidentally, Eqs.~(\ref{ec:Oalt}) make apparent the fact that $\mathcal{O}_{1,2}$ cancel when both the top and bottom quarks are on their mass shell, and they
also show that these two operators are not independent if either the top or the bottom quark is on-shell. For top on-shell,
\begin{equation}
\mathcal{O}_1 (\glp,\grp)  - \mathcal{O}_2 (\hl = \glp,\hr = \grp) = 0 \,,
\label{ec:tos}
\end{equation}
whereas for the bottom quark on-shell
\begin{equation}
\mathcal{O}_1 (\glp,\grp)  + \mathcal{O}_2 (\hl = \glp,\hr = \grp) = 0 \,,
\label{ec:bos}
\end{equation}
in obvious notation.
Here we will restrict ourselves to the cancellation among $\mathcal{O}_2$ and $\mathcal{O}_2^{(4)}$ for brevity, because the proof for $\mathcal{O}_1$ and $\mathcal{O}_1^{(4)}$ is almost identical, as it can be observed from their expressions in Eqs.~(\ref{ec:Oalt}).

\subsection{Cancellation in the $g b \to tW^-$ amplitude}

There are three Feynman diagrams contributing to this process, the two ones
in Fig.~\ref{fig:diag3}, involving a $Wtb$ vertex, and a third one with a quartic $gWtb$ coupling.
We denote by $p_1$, $p_2$, $p_3$ and $p_4$ the momenta of the external gluon, $b$ quark, top quark and $W$ boson, respectively.
The matrix element for the $t$-channel diagram reads
\begin{eqnarray}
\mathcal{M}_1 & = & \frac{g g_s}{\sqrt 2 M_W} \, \frac{1}{p_5^2 - m_t^2}\,
\bar u(p_3) \, \gamma^\nu (p_5\!\!\!\!\!\!\!\!\not\,\,\,\,\, +\, m_t)
\left[ (p_5\!\!\!\!\!\!\!\!\not\,\,\,\,\, -\, m_t) \gamma^\mu (\hr P_L + \hl P_R)  \right] u(p_2)
\nonumber \\[1mm]
& & \times \epsilon_\mu^* (p_4) \, \epsilon_\nu (p_1) \,,
\label{ec:M1}
\end{eqnarray}
using the expression for $\mathcal{O}_2$ in Eqs.~(\ref{ec:Oalt}) and the fact that the $b$ quark is on-shell, $p_2\!\!\!\!\!\!\!\not\,\,\,\, u(p_2) = m_b u(p_2)$.
The momentum of the internal top quark is $p_5=p_2-p_4$.
The factor between brackets corresponds to the $\mathcal{O}_2$ vertex, up to numerical constants. We have suppressed colour indices and a $\lambda^a/2$ factor, which are common to the three diagrams, and after colour averaging and summing amount to a global factor of $1/6$. 
The internal top quark propagator cancels with the $(p_t\!\!\!\!\!\!\!\not\,\,\,\, - m_t)$ vertex factor and the matrix element simplifies to
\begin{eqnarray}
\mathcal{M}_1 & = & \frac{g g_s}{\sqrt 2 M_W} \,
\bar u(p_3) \, \gamma^\nu \gamma^\mu (\hr P_L + \hl P_R) \, u(p_2)
\; \epsilon_\mu^* (p_4) \, \epsilon_\nu (p_1) \,.
\label{ec:M1s}
\end{eqnarray}
The matrix element for the $s$-channel diagram reads
\begin{eqnarray}
\mathcal{M}_2 & = & \frac{g g_s}{\sqrt 2 M_W} \, \frac{1}{p_6^2 - m_b^2}\,
\bar u(p_3)
\left[ (\hr P_L + \hl P_R) \gamma^\mu (p_6\!\!\!\!\!\!\!\not\,\,\,\, - m_b) \right] (p_6\!\!\!\!\!\!\!\not\,\,\,\, + m_b) \gamma^\nu
\, u(p_2) \nonumber \\[1mm]
& & \times \epsilon_\mu^* (p_4) \, \epsilon_\nu (p_1) \,,
\label{ec:M2}
\end{eqnarray}
where $p_6=p_1+p_2$ is the momentum of the internal $b$ quark, and we have used that
the external top quark is on-shell, $\bar u(p_3) p_3\!\!\!\!\!\!\!\not\,\,\,\,\,\,\, = \bar u(p_3) m_t$. This simplifies to
\begin{eqnarray}
\mathcal{M}_2 & = & \frac{g g_s}{\sqrt 2 M_W} \,
\bar u(p_3) \, \gamma^\mu \gamma^\nu (\hr P_L + \hl P_R) \, u(p_2)
\; \epsilon_\mu^* (p_4) \, \epsilon_\nu (p_1) \,.
\label{ec:M2s}
\end{eqnarray}
Finally, the diagram involving the quartic interaction gives
\begin{eqnarray}
\mathcal{M}_1^{(4)} & = & -\frac{\sqrt 2 g g_s}{M_W} \,
\bar u(p_3) \, g^{\mu \nu} (\hr P_L + \hl P_R) \, u(p_2)
\; \epsilon_\mu^* (p_4) \, \epsilon_\nu (p_1) \,,
\label{ec:M3}
\end{eqnarray}
and the cancellation follows from $\{\gamma^\mu,\gamma^\nu\} = 2 g^{\mu \nu}$.

\subsection{Cancellation in the $g g \to tW^- \bar b$ amplitude}

\begin{figure}[htb]
\begin{center}
\begin{tabular}{ccccc}
\epsfig{file=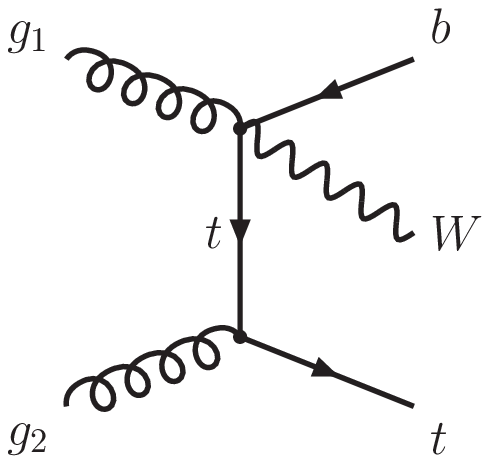,height=2.5cm,clip=} & \quad &
\epsfig{file=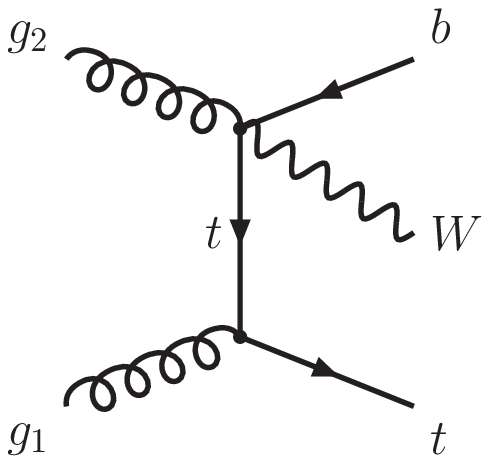,height=2.5cm,clip=} & \quad &
\epsfig{file=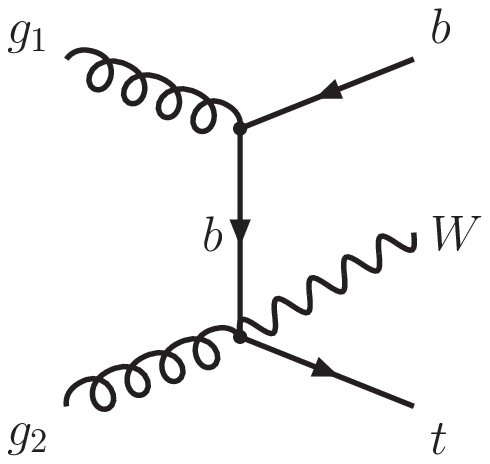,height=2.5cm,clip=} \\[2mm]
\epsfig{file=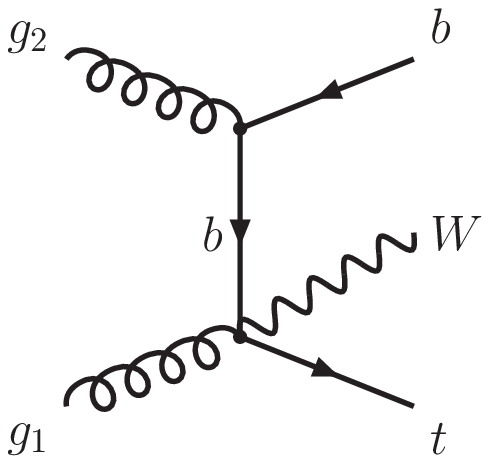,height=2.5cm,clip=} & &
\epsfig{file=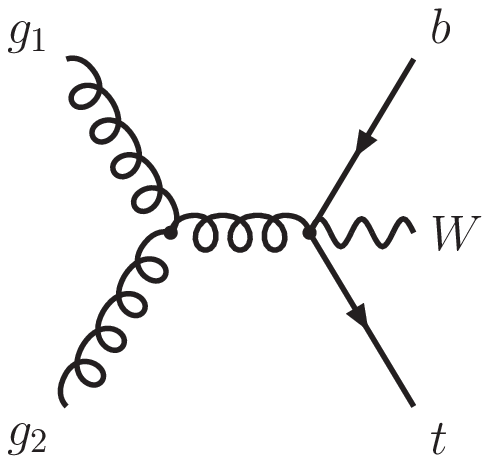,height=2.5cm,clip=}
\end{tabular}
\caption{Feynman diagrams for $gg \to tW^- \bar b$ involving anomalous quartic vertices.
The two initial state gluons are labelled as $g_1$ and $g_2$ for clarity.}
\label{fig:diag5}
\end{center}
\end{figure}

The diagrams with $Wtb$ vertices contributing to this process are the same as in the SM, shown in
Fig.~\ref{fig:diag4}, and the additional diagrams with quartic $gWtb$ vertices are displayed in Fig.~\ref{fig:diag5}. We number the diagrams from left to right and top to bottom.
There are two different colour flows contributing to the amplitude.
The two diagrams involving a triple gluon vertex contribute to both colour flows but
cancel with the last diagram in Fig.~\ref{fig:diag5},
\begin{equation}
\mathcal{M}_5 + \mathcal{M}_8 + \mathcal{M}_3^{(4)} = 0\,,
\end{equation}
in the same way
as for $gb \to tW^-$, explained in the previous subsection. Then, we concentrate ourselves on the remaining diagrams. The momenta of the two initial state gluons are $p_1$ and $p_2$, and the momenta of the top quark, $W$ boson and $b$ quark are $p_3$, $p_4$ and $p_5$, respectively. Inserting the expression of $\mathcal{O}_2$ from Eqs.~(\ref{ec:Oalt}), using the equations of motion for the external top and bottom quarks and simplifying the propagators, the matrix elements for the remaining diagrams are
\begin{align}
\mathcal{M}_1 & = \frac{g g_s^2}{\sqrt 2 M_W} \, \left[
\frac{1}{p_6^2-m_t^2} \, \bar u(p_3) \, \gamma^\nu (p_6\!\!\!\!\!\!\!\not\,\,\,\, + m_t)
   \gamma^\mu \gamma^\sigma (\hr P_L + \hl P_R) \, v(p_5) \right. \nonumber \\
& \left. + \frac{1}{p_7^2-m_b^2} \, \bar u(p_3) \, (\hr P_L + \hl P_R)
   \gamma^\nu \gamma^\mu (p_7\!\!\!\!\!\!\!\not\,\,\,\, + m_b) \gamma^\sigma \, v(p_5)
  \right] \times \epsilon \,, \nonumber \\[1mm]
\displaybreak
\mathcal{M}_2 & = \frac{g g_s^2}{\sqrt 2 M_W} \, \left[
\frac{1}{p_9^2-m_t^2}\, \bar u(p_3) \, \gamma^\sigma (p_9\!\!\!\!\!\!\!\not\,\,\,\, + m_t)
   \gamma^\mu \gamma^\nu (\hr P_L + \hl P_R) \, v(p_5) \right. \nonumber \\
& \left. + \frac{1}{p_{10}^2-m_b^2} \, \bar u(p_3) \, (\hr P_L + \hl P_R)
  \gamma^\sigma \gamma^\mu (p_{10}\!\!\!\!\!\!\!\!\!\not\,\,\,\,\,\, + m_b) \gamma^\nu \,
   v(p_5) \right] \times \epsilon \,, \nonumber \\[1mm]
%
\mathcal{M}_3 & = \frac{g g_s^2}{\sqrt 2 M_W} \,
\frac{1}{p_7^2-m_b^2} \, \bar u(p_3) \, (\hr P_L + \hl P_R) \gamma^\mu \gamma^\nu 
(p_7\!\!\!\!\!\!\!\not\,\,\,\, + m_b)  \gamma^\sigma \, v(p_5) 
\times \epsilon \,, \nonumber \\
\mathcal{M}_4 & = \frac{g g_s^2}{\sqrt 2 M_W} \,
\frac{1}{p_{10}^2-m_b^2} \, \bar u(p_3) \, (\hr P_L + \hl P_R) \gamma^\mu \gamma^\sigma 
(p_{10}\!\!\!\!\!\!\!\!\!\not\,\,\,\,\,\, + m_b)  \gamma^\nu \, v(p_5) 
\times \epsilon \,, \nonumber \\
\mathcal{M}_6 & = \frac{g g_s^2}{\sqrt 2 M_W} \,
\frac{1}{p_9^2-m_t^2} \, \bar u(p_3) \, \gamma^\sigma (p_9\!\!\!\!\!\!\!\not\,\,\,\, + m_t)
\gamma^\nu \gamma^\mu (\hr P_L + \hl P_R) \, v(p_5) 
\times \epsilon \,, \nonumber \\
\mathcal{M}_7 & = \frac{g g_s^2}{\sqrt 2 M_W} \,
\frac{1}{p_6^2-m_t^2} \, \bar u(p_3) \, \gamma^\nu (p_6\!\!\!\!\!\!\!\not\,\,\,\, + m_t)
\gamma^\sigma \gamma^\mu (\hr P_L + \hl P_R) \, v(p_5) 
\times \epsilon \,,
\end{align}
where $\epsilon$ stands for $\epsilon_\mu^*(p_4) \epsilon_\nu(p_2) \epsilon_\sigma(p_1)$.
The additional momenta introduced are $p_6 = p_3-p_2$, $p_7 = p_1-p_5$, $p_9 = p_3-p_1$,
$p_{10} = p_2-p_5$. The first four diagrams in Fig.~\ref{fig:diag5} give
\begin{eqnarray}
\mathcal{M}_1^{(4)} & = & -\frac{\sqrt 2 g g_s^2}{M_W} \, \frac{1}{p_6^2-m_t^2}
\bar u(p_3) \, \gamma^\nu (p_6\!\!\!\!\!\!\!\not\,\,\,\, + m_t) g^{\mu \sigma}
(\hr P_L + \hl P_R) \, v(p_5) \times \epsilon \,, \nonumber \\
\mathcal{M}_2^{(4)} & = & -\frac{\sqrt 2 g g_s^2}{M_W} \, \frac{1}{p_9^2-m_t^2}
\bar u(p_3) \, \gamma^\sigma (p_9\!\!\!\!\!\!\!\not\,\,\,\, + m_t) g^{\mu \nu}
(\hr P_L + \hl P_R) \, v(p_5) \times \epsilon \,, \nonumber \\
\mathcal{M}_3^{(4)} & = & -\frac{\sqrt 2 g g_s^2}{M_W} \, \frac{1}{p_7^2-m_b^2}
\bar u(p_3) \, (\hr P_L + \hl P_R) g^{\mu \nu} (p_7\!\!\!\!\!\!\!\not\,\,\,\, + m_b) 
\gamma^\sigma \, v(p_5) \times \epsilon \,, \nonumber \\
\mathcal{M}_4^{(4)} & = & -\frac{\sqrt 2 g g_s^2}{M_W} \, \frac{1}{p_{10}^2-m_b^2}
\bar u(p_3) \, (\hr P_L + \hl P_R) g^{\mu \sigma}
(p_{10}\!\!\!\!\!\!\!\!\!\not\,\,\,\,\,\, + m_b) 
\gamma^\nu \, v(p_5) \times \epsilon \,.
\end{eqnarray}
Using $\{\gamma^\mu,\gamma^\nu\} = 2 g^{\mu \nu}$,
for one colour flow we have
\begin{equation}
\mathcal{M}_1 + \mathcal{M}_3 + \mathcal{M}_7 + \mathcal{M}_1^{(4)}
+ \mathcal{M}_3^{(4)} = 0 \,,
\end{equation}
and for the other one
\begin{equation}
\mathcal{M}_2 + \mathcal{M}_4 + \mathcal{M}_6 + \mathcal{M}_2^{(4)}
+ \mathcal{M}_4^{(4)} = 0 \,.
\end{equation}
Finally, we point out that introducing a gauge non-invariant regulator for the top quark pole singularities, like the usual prescription for the propagators $p^2 - m_t^2 \to p^2 - m_t^2 + i m_t \Gamma_t$, spoils the cancellation. Alternative gauge-invariant prescriptions can be found for example in Ref.~\cite{Kauer:2001sp}.

\section{Top decay observables}
\label{sec:b}

For completeness, we briefly introduce here the definitions of top decay observables used in the combination with single top cross sections. An extended discussion regarding these observables
can be found in Ref.~\cite{AguilarSaavedra:2006fy}. 

The charged lepton angular distribution in the $W$ rest frame is very sensitive
to $Wtb$ anomalous couplings.
It is defined in terms of the angle $\thlw$ between the charged lepton in the $W$ rest frame and the $W$ momentum in the top quark rest frame.
An important point is that this distribution is independent of the top quark production mechanism, in particular independent of the top polarisation, and thus
the related observables can be defined and measured either in top pair or single top production.
The normalised angular distribution of the charged lepton can be written as
\begin{equation}
\frac{1}{\Gamma} \frac{d \Gamma}{d \cos \thlw} = \frac{3}{8}
(1 + \cos \thlw)^2 \, \fp + \frac{3}{8} (1-\cos \thlw)^2 \, \fm
+ \frac{3}{4} \sin^2 \thlw \, \fz \,.
\label{ec:dist}
\end{equation}
The helicity fractions $F_i \equiv \Gamma_i / \Gamma$ are the normalised partial widths for the top decay to the three $W$ helicity states (we drop here the $t$ subindex in the top width for simplicity).
A fit to the
$\cos \thlw$ distribution allows to extract from experiment the values of $\Fi$,
which are not independent but satisfy $\fp + \fm + \fz = 1$ by definition.
From the measurement of helicity fractions one can constrain $Wtb$ anomalous couplings \cite{Kane:1991bg}. Alternatively, from this
distribution one can measure the helicity ratios
\begin{equation}
\rhpm \equiv \frac{\Gpm}{\Gz} = \frac{\Fpm}{\fz} \,,
\label{ec:rho}
\end{equation}
which are independent quantities. For any fixed $z$ in the interval $[-1,1]$, one can also define an asymmetry 
\begin{equation}
A_z = \frac{N(\cos \thlw > z) - N(\cos \thlw < z)}{N(\cos \thlw > z) +
N(\cos \thlw < z)} \,.
\end{equation}
The most obvious choice is $z=0$, giving the forward-backward asymmetry
$\afb$ \cite{Lampe:1995xb,delAguila:2002nf}.
But more convenient choices are $z = \mp (2^{2/3}-1)$ \cite{AguilarSaavedra:2006fy}. Defining
$\beta = 2^{1/3}-1$, we have
\begin{eqnarray}
z = -(2^{2/3}-1) & \rightarrow & A_z = \Ap = 3 \beta [\fz+(1+\beta) \fp] \,,
\notag \\
z = (2^{2/3}-1) & \rightarrow & A_z = \Am = -3 \beta [\fz+(1+\beta) \fm] \,.
\label{ec:apm}
\end{eqnarray}
These asymmetries are obviously determined by the $W$ helicity fractions (or ratios) and conversely, from their measurement the helicity fractions and ratios can be reconstructed.
Moreover, the charged lepton energy distribution in the top quark rest frame is uniquely determined by the helicity fractions and the top, bottom and $W$ boson masses. Among these observables we select $\Apm$ and $\rhpm$, which are the most sensitive to $Wtb$ anomalous couplings, for the combination with single top cross sections. They take the tree-level SM values
$\Ap = 0.5482$, $\Am = -0.8397$, $\rhp = 5.1 \times 10^{-4}$ and $\rhm = 0.423$, for a top quark on its mass shell.

Further observables can be built involving the top spin.
For the decay $t \to W^+ b \to \ell^+ \nu b,q \bar q' b$, the angular
distributions of $X=\ell^+,\nu,q,\bar q',W^+,b$ (which are
called ``spin analysers'') in the top quark rest frame are given by
\cite{Jezabek:1994zv,Jezabek:1994qs}
\begin{equation}
\frac{1}{\Gamma} \frac{d\Gamma}{d \cos \theta_X} = \frac{1}{2} (1+\hk_X \cos
\theta_X )
\label{ec:tdist}
\end{equation}
with $\theta_X$ the angle between the three-momentum of $X$ in the $t$
rest frame and the top spin direction.
The constants $\hk_X$  are called ``spin analysing power'' of $X$ and can range
between $-1$ and $1$.
For the decay of a top antiquark the distributions are the same, with
$\hk_{\bar X} = - \hk_X$ as long as CP is conserved in the decay. 
In the SM,
$\hk_{\ell^+} = \hk_{\bar q'} = 1$,
$\hk_\nu = \hk_q = -0.319$, $\hk_{W^+} = - \hk_b = 0.406$ at the tree level
($q$ and $q'$ are the up- and down-type quarks, respectively,
resulting from the $W$ decay).
Tree-level expressions of the spin analysing power constants
for a CP-conserving $Wtb$ vertex with $t$, $b$ on-shell as in Eq.~(\ref{ec:1})
have been obtained in Ref.~\cite{AguilarSaavedra:2006fy} within the narrow
width approximation.

In the $t$-channel single top process the top quarks are produced with a high degree of polarisation
along the direction of the final state jet \cite{Mahlon:1999gz}. The
corresponding distributions are
\begin{equation}
\frac{1}{\Gamma} \frac{d\Gamma}{d \cos \theta_{X}} = \frac{1}{2} (1+ P \hk_X \cos
\theta_{X} ) \,,
\end{equation}
where the angles $\theta_X$ are now measured using as spin direction the jet
three-momentum in the top quark rest frame, and $P\simeq 0.89$ is the top polarisation along this axis. (This value, calculated with the matching of $tj/t \bar b j$ at 10 GeV, is in good agreement with the polarisation calculated 
with the subtraction method \cite{Mahlon:1999gz}.)
Forward-backward asymmetries can be built using these distributions,
\begin{equation}
A_{X} = \frac{N(\cos \theta_X > 0) - N(\cos \theta_X > 0)}
{N(\cos \theta_X > 0) + N(\cos \theta_X > 0)} = 
\frac{1}{2} P \hk_X \,.
\end{equation}
These spin asymmetries depend on the top polarisation $P$ but their ratios do not, for example
\begin{eqnarray}
\frac{A_b}{A_\ell} & = & \frac{\hk_b}{\hk_\ell} \equiv \rbl \,, \notag \\
\frac{A_\nu}{A_\ell} & = & \frac{\hk_\nu}{\hk_\ell} \equiv \rnl
\end{eqnarray}
only depend on anomalous couplings and the masses involved. Additionally, their experimental systematic errors are expected to be smaller than for the spin asymmetries themselves. As it happens with $\Apm$ and $\rhpm$, these asymmetry ratios can be measured in top pair production as well.
In $t \bar t$ production the top quarks are unpolarised at the tree level, but
their spins are correlated \cite{Stelzer:1995gc,Bernreuther:2004jv}. Suitable ratios of spin correlation asymmetries can provide a measurement of $\rbl$ and $\rnl$ but the precision will likely be worse than in single top production because
spin correlation asymmetries are numerically smaller, {\em e.g.} by a factor of 8 in the $t \bar t$ semileptonic decay channel.

\section{Effect of top quark decay}
\label{sec:c}

Here we examine the effect of top decay in the $\kappa$ factors,
concentrating on the $t$-channel process for brevity.
 We calculate the $\kappa$ factors for the processes with top decay numerically, following the procedure explained in section~\ref{sec:4} for $t W^- \bar b$ production. For the processes with top on-shell the calculation is carried out as in section \ref{sec:2}, with the difference that here for the sake of computational speed we perform the calculations for $tj$ without the veto on $p_T^{\bar b_\text{ISR}}$, which is sufficient to illustrate the differences. Results for the quadratic terms are shown in Table~\ref{tab:koff}.
As expected, the effect of the top width is small, at the $1-2\%$ level, and in most cases it is smaller than the theoretical uncertainty. The differences have little numerical relevance in the limits obtained for anomalous couplings.

\begin{table}[ht]
\begin{center}
\begin{small}
\begin{tabular}{ccccccccccccc}
& \quad 
& \multicolumn{2}{c}{$tj$} & \quad & \multicolumn{2}{c}{$t \bar b j$} & \quad
& \multicolumn{2}{c}{$\bar tj$} & \quad & \multicolumn{2}{c}{$\bar t b j$} \\
&& on & off && on & off && on & off && on & off \\
$\vr^2$ && 0.895 & 0.888 && 0.927 & 0.917 && 1.101 & 1.093 && 1.068 & 1.064 \\
$\gl^2$ && 1.47 & 1.47 && 1.96 & 1.96 && 2.12 & 2.13 && 2.98 & 3.01 \\
$\gr^2$ && 2.00 & 1.98 && 2.97 & 2.95 && 1.63 & 1.62 && 2.08 & 2.07
\end{tabular}
\end{small}
\end{center}
\caption{$\kappa$ factors for the quadratic terms in the $tj$, $t \bar b j$ processes
and their charge conjugate, calculated for top on-shell or decaying $t \to W^+ b \to \ell \nu b$ (labelled as ``on'' and ``off'', respectively).}
\label{tab:koff}
\end{table}

\end{document}